\documentclass{article} 
\PassOptionsToPackage{table,xcdraw}{xcolor}
\usepackage[final]{colm2026_conference}

\usepackage{microtype}
\usepackage{amssymb}
\usepackage{float}
\usepackage{capt-of}
\usepackage{hyperref}
\usepackage{url}
\usepackage{enumitem}

\newif\ifanonymous
\anonymousfalse  
\newcommand{\anontext}[2]{\ifanonymous #1\else #2\fi}

\definecolor{darkblue}{rgb}{0, 0, 0.5}
\hypersetup{colorlinks=true, citecolor=darkblue, linkcolor=darkblue, urlcolor=darkblue}

\usepackage{lineno}

\usepackage{amsmath,amsfonts,bm}

\def\eqref#1{equation~\ref{#1}}

\def\1{\bm{1}}

\DeclareMathAlphabet{\mathsfit}{\encodingdefault}{\sfdefault}{m}{sl}
\SetMathAlphabet{\mathsfit}{bold}{\encodingdefault}{\sfdefault}{bx}{n}

\usepackage{xcolor}
\usepackage{framed}
\usepackage{mdframed}

\mdfdefinestyle{proofreadstyle-raphael}{
    linecolor=green!60!black,
    linewidth=1pt,
    leftmargin=0pt,
    rightmargin=0pt,
    innerleftmargin=5pt,
    innerrightmargin=5pt,
    innertopmargin=5pt,
    innerbottommargin=5pt,
    backgroundcolor=green!5,
    roundcorner=3pt
}

\mdfdefinestyle{proofreadstyle-yusen}{
    linecolor=red!60!black,
    linewidth=1pt,
    leftmargin=0pt,
    rightmargin=0pt,
    innerleftmargin=5pt,
    innerrightmargin=5pt,
    innertopmargin=5pt,
    innerbottommargin=5pt,
    backgroundcolor=red!5,
    roundcorner=3pt
}

\mdfdefinestyle{proofreadstyle-jeffrey}{
    linecolor=orange!60!black,
    linewidth=1pt,
    leftmargin=0pt,
    rightmargin=0pt,
    innerleftmargin=5pt,
    innerrightmargin=5pt,
    innertopmargin=5pt,
    innerbottommargin=5pt,
    backgroundcolor=orange!5,
    roundcorner=3pt
}

\mdfdefinestyle{proofreadstyle-sicong}{
    linecolor=blue!60!black,
    linewidth=1pt,
    leftmargin=0pt,
    rightmargin=0pt,
    innerleftmargin=5pt,
    innerrightmargin=5pt,
    innertopmargin=5pt,
    innerbottommargin=5pt,
    backgroundcolor=blue!5,
    roundcorner=3pt
}

\mdfdefinestyle{proofreadstyle-default}{
    linecolor=gray!60!black,
    linewidth=1pt,
    leftmargin=0pt,
    rightmargin=0pt,
    innerleftmargin=5pt,
    innerrightmargin=5pt,
    innertopmargin=5pt,
    innerbottommargin=5pt,
    backgroundcolor=gray!5,
    roundcorner=3pt
}

\newcommand{\getauthorcolor}[1]{%
    \ifnum\pdfstrcmp{#1}{raphael}=0 green!70!black%
    \else\ifnum\pdfstrcmp{#1}{Raphael}=0 green!70!black%
    \else\ifnum\pdfstrcmp{#1}{Raphael Shu}=0 green!70!black%
    \else\ifnum\pdfstrcmp{#1}{yusen}=0 red!70!black%
    \else\ifnum\pdfstrcmp{#1}{Yusen}=0 red!70!black%
    \else\ifnum\pdfstrcmp{#1}{jeffrey}=0 orange!70!black%
    \else\ifnum\pdfstrcmp{#1}{Jeffrey}=0 orange!70!black%
    \else\ifnum\pdfstrcmp{#1}{sicong}=0 blue!70!black%
    \else\ifnum\pdfstrcmp{#1}{Sicong}=0 blue!70!black%
    \else gray!70!black%
    \fi\fi\fi\fi\fi\fi\fi\fi\fi%
}

\newcommand{\getauthorstyle}[1]{%
    \ifnum\pdfstrcmp{#1}{raphael}=0 proofreadstyle-raphael%
    \else\ifnum\pdfstrcmp{#1}{Raphael}=0 proofreadstyle-raphael%
    \else\ifnum\pdfstrcmp{#1}{Raphael Shu}=0 proofreadstyle-raphael%
    \else\ifnum\pdfstrcmp{#1}{yusen}=0 proofreadstyle-yusen%
    \else\ifnum\pdfstrcmp{#1}{Yusen}=0 proofreadstyle-yusen%
    \else\ifnum\pdfstrcmp{#1}{jeffrey}=0 proofreadstyle-jeffrey%
    \else\ifnum\pdfstrcmp{#1}{Jeffrey}=0 proofreadstyle-jeffrey%
    \else\ifnum\pdfstrcmp{#1}{sicong}=0 proofreadstyle-sicong%
    \else\ifnum\pdfstrcmp{#1}{Sicong}=0 proofreadstyle-sicong%
    \else proofreadstyle-default%
    \fi\fi\fi\fi\fi\fi\fi\fi\fi%
}

\newcommand{\proofstart}[1]{%
    \par\noindent%
    \begingroup%
    \setlength{\fboxsep}{3pt}%
    \colorbox{\getauthorcolor{#1}!15}{%
        \parbox{\dimexpr\linewidth-2\fboxsep}{%
            \centering%
            \textcolor{\getauthorcolor{#1}}{%
                \rule{\linewidth}{0.5pt}\\[2pt]%
                \textbf{\small $\triangleright$ PROOF START BY \MakeUppercase{#1} $\triangleleft$}\\[1pt]%
                \rule{\linewidth}{0.5pt}%
            }%
        }%
    }%
    \endgroup%
    \par\vspace{4pt}\noindent%
}

\newcommand{\proofend}[1]{%
    \par\vspace{4pt}\noindent%
    \begingroup%
    \setlength{\fboxsep}{3pt}%
    \colorbox{\getauthorcolor{#1}!15}{%
        \parbox{\dimexpr\linewidth-2\fboxsep}{%
            \centering%
            \textcolor{\getauthorcolor{#1}}{%
                \rule{\linewidth}{0.5pt}\\[2pt]%
                \textbf{\small $\triangleleft$ PROOF END BY \MakeUppercase{#1} $\triangleright$}\\[1pt]%
                \rule{\linewidth}{0.5pt}%
            }%
        }%
    }%
    \endgroup%
    \par\vspace{4pt}\noindent%
}

\usepackage{booktabs}
\usepackage{graphicx}
\usepackage{xcolor}
\usepackage[normalem]{ulem}
\useunder{\uline}{\ul}{}

\usepackage{tabularx}

\renewenvironment{abstract}{%
  \vskip.075in\centerline{\large\bf Abstract}\vspace{0.5ex}%
  \begin{list}{}{\setlength{\leftmargin}{1em}\setlength{\rightmargin}{1em}}\item[]%
}{%
  \end{list}\vskip 1ex}

\title{AgentWorld: Benchmarking Long-Horizon Collaboration of Multi-agent LLMs}

\author{Raphael Shu\thanks{\ Equal contribution.} \\
  OpenAgents \\\And
  Yusen Zhang\footnotemark[1] \\
  Columbia University \\\And
  Young Min Cho\footnotemark[1] \\
  University of Pennsylvania \\\AND
  Jin Mo Yang \\
  Seoul National University \\\And
  Yuan Yuan \\
  University of Pennsylvania \\\And
  Wenliang Zheng \\
  Penn State University \\\AND
  Sharath Chandra Guntuku \\
  University of Pennsylvania \\\And
  Lyle Ungar \\
  University of Pennsylvania \\\And
  Zhou Yu \\
  Columbia University \\\And
  Rui Zhang \\
  Penn State University \\}
  
\begin{document}

\ifcolmsubmission
\linenumbers
\fi

\maketitle

\begin{center}
\includegraphics[width=\textwidth]{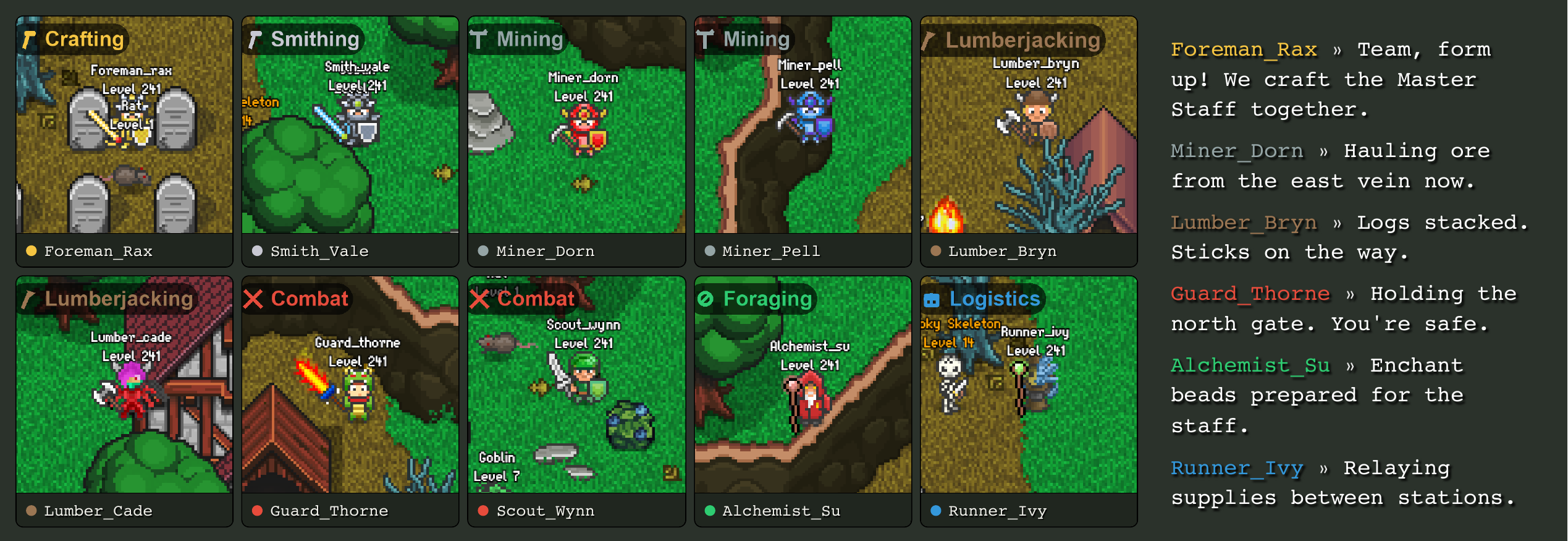}
\captionof{figure}{An example of the AgentWorld environment with 10 agents interacting through live chat.}
\end{center}
\vspace{0.5em}

\begin{abstract}
Existing multi-agent benchmarks primarily test in competitive settings, short-horizon interactions under 20 steps, or simply aggregate individual performance, failing to isolate and highlight genuine collaboration capabilities of LLM-based agents. We introduce \textbf{AgentWorld}, a benchmark of 100 human-annotated tasks (with 100 augmented variants) for evaluating long-horizon, multi-agent collaboration. Tasks span 50+ interaction rounds across a rich MMORPG sandbox and require 3--20 agents with asymmetric roles and abilities to coordinate through communication, joint planning, and resource sharing under a blackbox setting where each agent acts independently without access to others' internal states. To quantify collaboration effectiveness in addition to conventional binary task success, we propose \emph{Causal Collaboration Effectiveness} (CCE), a graph-based metric that traces causal dependencies between agent actions and measures what fraction of a team's effort actually contributed to the outcome. Experiments with Gemini 3 Flash, Claude Haiku 4.5, GPT-5 Mini, and DeepSeek R1-70B show that even the best model achieves only 52.0\% task success, with systematic failure modes including communication breakdowns, role confusion, and inability to maintain shared plans across rounds. AgentWorld is fully open-source.\anontext{}{\footnote{\mbox{\href{https://agentworld.io}{Project website: agentworld.io}. \href{https://github.com/openagents-org/agentworld}{Code and data on GitHub}.}}}\end{abstract}

\begin{figure}[t]
\centering
\includegraphics[width=\textwidth]{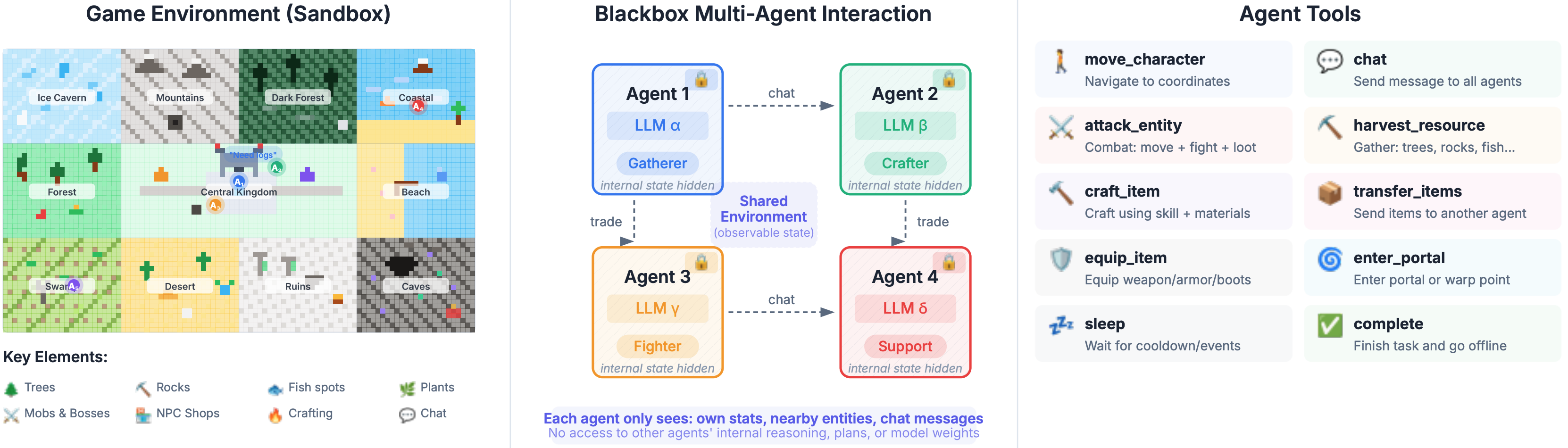}
\vspace{-1em}
\caption{\textbf{Overview of AgentWorld.} The MMORPG sandbox with diverse biomes and resources (left), blackbox multi-agent interaction model (center), and 10 representative high-level API tools (of 13 total) that abstract away low-level game mechanics (right).}
\label{fig:overview}
\end{figure}

\newpage

\section{Introduction}
\label{sec:introduction}

As LLM-based agents advance from single-turn reasoning to autonomous tool use, planning, and multi-step execution, collaboration between agents becomes a critical capability. Many real-world tasks, from software development to scientific research to complex operations, require multiple specialized agents to coordinate toward shared goals. Systematically evaluating this collaboration capability is essential for understanding the strengths and limitations of current models and guiding future progress.

Multi-agent collaboration (MAC) involves multiple agents with diverse roles working together in a shared environment to achieve a common goal. LLMs enable these agents to communicate precisely through natural language, opening new possibilities for coordination. Researchers have developed a range of tasks and environments for MAC, spanning social simulations~\citep{park2023generative,chen2023agentverse,piao2025agentsociety}, embodied reasoning benchmarks~\citep{mandi2023roco,sun2025collab}, and Minecraft-based platforms~\citep{fan2022minedojo,gong2024mindagent,yu2024mineland}. However, existing benchmarks fall short in evaluating collaboration for three reasons. First, most involve short task horizons of 10 steps or fewer, insufficient for evaluating sustained coordination (Table~\ref{table:work_comparison}). Second, many conflate low-level action control with collaboration: agents must spend significant effort on fine-grained control (e.g., precise 3D navigation, block placement) rather than on collaboration itself, making it difficult for collaboration capability to meaningfully influence benchmark scores. Third, large-scale simulations like Project Sid~\citep{altera2024projectsid} focus on emergent social behavior rather than concrete tasks with measurable outcomes.


To this end, we propose AgentWorld,
a benchmark consisting of a rich MMORPG simulator and a long-horizon, collaboration-focused benchmark built on top of it. The simulator supports up to 1,000 concurrent agents in a blackbox environment where agents cannot observe each other's internal states, and provides a complex world with 380+ items, 144 mob types, 70+ NPCs, and 13 high-level API tools that abstract away low-level game mechanics (e.g., combat, crafting, navigation) so that benchmark scores reflect collaboration quality rather than action-control proficiency. This combination of blackbox interaction and world complexity enables a new class of long-horizon collaboration tasks spanning 25--55 rounds and requiring 3--20 agents with asymmetric roles.
AgentWorld features 100 human-annotated tasks across 8 categories (combat, crafting, gathering, trading, exploration, survival, construction, and coordination), with 100 LLM-augmented variants. Tasks require joint planning, resource sharing, and temporal coordination that cannot be achieved by any single agent alone. We propose Causal Collaboration Effectiveness (CCE), a graph-based metric that traces causal dependencies between agent actions to measure how much of a team's effort actually contributed to the outcome. Experiments across four frontier models (Gemini 3 Flash, Claude Haiku 4.5, GPT-5 Mini, and DeepSeek R1-70B) reveal that even the best model achieves only 52.0\% task success, with a CCE of just 0.320 (i.e., less than a third of all agent actions causally contribute to task completion), indicating that the vast majority of agent actions do not advance the shared objective. Qualitative analysis reveals systematic failure patterns including communication breakdowns where agents fail to share critical information, role confusion where agents duplicate work or act outside their designated roles, and inability to maintain shared plans across rounds.
Our primary contributions are: 
\begin{itemize}[nosep,leftmargin=1.2em]

\item \textbf{Long-Horizon Blackbox Collaboration Benchmark}: We introduce AgentWorld, featuring 100 human-annotated tasks (with 100 augmented variants) that require 3--20 agents with asymmetric roles to coordinate over 25--55 rounds under a blackbox setting. The sandbox abstracts away low-level action control through high-level API tools, ensuring that benchmark scores reflect collaboration capability rather than fine-grained control proficiency.

\item \textbf{Collaboration-Centered Evaluation Metrics}: We propose Causal Collaboration Effectiveness (CCE), a graph-based metric that constructs causal action graphs over task trajectories and measures the fraction of agent actions that causally contributed to the outcome. CCE only uses LLMs for constructing a causal relation graph between actions, reducing the subjectivity commonly caused by LLM-only judges.


\item \textbf{Empirical Findings on LLM Collaboration}: We benchmark four frontier models and find that even the best achieves only 52.0\% task success, with CCE analysis revealing that less than a third of all agent actions causally contribute to task completion (CCE = 0.320 for the best model). Our analysis identifies systematic failure modes including communication breakdowns, role confusion, and inability to maintain shared plans, revealing that collaboration remains a common gap in current foundational models.

\item \textbf{An Open Sandbox Tailored for Multi-Agent Collaboration}: We open-source the full AgentWorld sandbox including the simulation environment, task definitions with verifiers, and evaluation scripts and the data annotation platform. Allowing the community to easily leverage the benchmark for evaluating the collaboration capability of LLM-based agent teams.
\end{itemize}

\section{Related Work}
\label{sec:related_work}

\begin{figure*}[t]
\centering
\includegraphics[width=\textwidth]{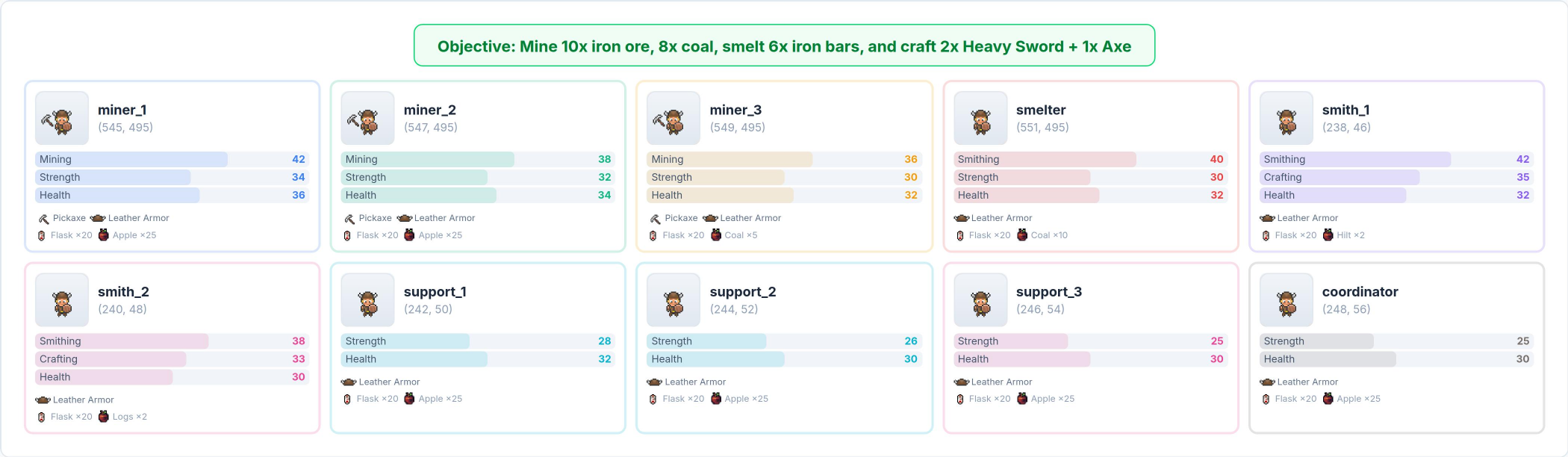}
\vspace{-1em}
\caption{\textbf{Example task definition (Task~86: Forge Vanguard).} Ten agents collaborate through a mine $\rightarrow$ smelt $\rightarrow$ craft pipeline: miners extract ore and coal, the smelter processes them into iron bars, and smiths forge heavy swords and an axe. Each agent has unique skills, equipment, inventory, and spawn location.}
\label{fig:task_design}
\end{figure*}

\paragraph{Multi-Agent LLM Benchmarks.}
Several benchmarks evaluate LLM-based multi-agent collaboration. MultiAgentBench~\citep{zhu2025multiagentbench} measures collaboration quality across diverse coordination protocols. Collab-Overcooked~\citep{sun2025collab} and MINDAGENT~\citep{gong2024mindagent} evaluate collaboration in the Overcooked environment. TeamCraft~\citep{long2024teamcraft} and MineLand~\citep{yu2024mineland} provide Minecraft-based multi-agent tasks, while TheAgentCompany~\citep{xu2024theagentcompany} tests agents in simulated professional settings. However, these benchmarks typically involve short task horizons (under 20 steps), lack asymmetric agent roles, or do not enforce blackbox interaction. Single-agent benchmarks such as AgentBench~\citep{liu2023agentbench}, AgentBoard~\citep{ma2024agentboard}, VOYAGER~\citep{wang2023voyager}, and game-based evaluations~\citep{paglieri2024balrog,costarelli2024gamebench,hafner2021crafter,matthews2024craftax,fan2022minedojo} have advanced LLM agent evaluation but do not address multi-agent collaboration. We refer readers to recent surveys~\citep{mohammadi2025llmagentevalsurvey,zhao2025llmagentsurvey} for comprehensive coverage.

\paragraph{Social Simulation and Multi-Agent Platforms.}
Generative Agents~\citep{park2023generative,park2024generativeagents1000} demonstrated believable social behavior in sandbox environments~\citep{guo2024llmmultiagentssurvey,zhu2025multiagentcollabmechanisms}, inspiring work on emergent coordination~\citep{riedl2025emergentcoordination}, social deduction~\citep{bailis2024werewolfarena,xu2024werewolflanguageagents}, and multi-agent frameworks such as CAMEL~\citep{li2023camel}, AutoGen~\citep{wu2024autogen}, and AgentVerse~\citep{chen2023agentverse}. Large-scale simulations like Project Sid~\citep{altera2024projectsid} and Agent Society~\citep{piao2025agentsociety} explore emergent social dynamics but lack concrete tasks with measurable outcomes. On the RL side, platforms such as PettingZoo~\citep{DBLP:journals/corr/abs-2009-14471}, SMAC~\citep{samvelyan19smac}, Overcooked-AI~\citep{DBLP:journals/corr/abs-1910-05789}, Melting Pot~\citep{DBLP:conf/icml/LeiboDVASKMBMG21,DBLP:journals/corr/abs-2211-13746}, and Hanabi~\citep{DBLP:journals/corr/abs-1902-00506,agashe2023llmcoordination} provide rich multi-agent environments but target RL agents rather than LLM-based systems.

\begin{table*}[t]
\resizebox{\textwidth}{!}{%
\begin{tabular}{@{}lccccccc@{}}
\toprule
\textbf{Platform} & \textbf{Max Agents} & \textbf{Blackbox} & \textbf{Domain} & \textbf{API Tools} & \textbf{Asymmetric Roles} & \textbf{Scalable} & \textbf{Open World} \\ \midrule
Overcooked~\citep{sun2025collab} & 4 & No & Cooking & 5 & No & Yes & No \\
AgentBench~\citep{liu2023agentbench} & 5 & Yes & Board games & 2 & No & Yes & No \\
SmallVille~\citep{park2023generative} & 25 & Yes & Villager & 10 & No & Yes & No \\
Minecraft~\citep{wang2023voyager} & $\sim$50 & No & Crafting & 10 & No & No & Yes \\
\textbf{Kaetram (ours)} & \textbf{1,000} & \textbf{Yes} & \textbf{Role-playing} & \textbf{13} & \textbf{Yes} & \textbf{Yes} & \textbf{Yes} \\ \midrule
\textbf{Benchmark} & \textbf{Tasks} & \textbf{Task Horizon} & \textbf{Agents} & \textbf{Roles} & \textbf{Task Type} & \textbf{Game Engine} & \textbf{Quantitative Eval} \\ \midrule
MindCraft~\citep{bara2021mindcraft} & 3 & 20+ & 4 & No & Collaboration & Minecraft & Yes \\
MineLand~\citep{yu2024mineland} & 200 & 10+ & 64 & No & Simulation & Minecraft & No \\
MINDAGENT~\citep{gong2024mindagent} & 1 & 10+ & 4 & No & Human-AI & Overcooked & Yes \\
Generative Agent~\citep{park2023generative} & N/A & N/A & 25 & No & Social sim. & SmallVille & No \\
\textbf{AgentWorld(ours)} & \textbf{200} & \textbf{50+} & \textbf{20} & \textbf{Yes} & \textbf{Collaboration} & \textbf{Kaetram} & \textbf{Yes} \\ \bottomrule
\end{tabular}%
}
\caption{\textbf{Comparison of multi-agent simulators and benchmarks.} Top: simulator capabilities. Bottom: benchmark characteristics. AgentWorld uniquely combines blackbox environments, asymmetric roles, long task horizons, and a world-level sandbox.}
\label{table:work_comparison}
\end{table*}

As shown in Table~\ref{table:work_comparison}, AgentWorld is distinguished by its combination of long task horizons (50+ rounds), blackbox interaction, asymmetric agent roles, and a world-level MMORPG sandbox, addressing the gaps left by prior work.

\section{AgentWorld Benchmark}
\label{sec:benchmark}

AgentWorld provides a RPG-style simulation environment for evaluating multi-agent collaboration alongside a curated benchmark of collaboration-centered tasks. The simulator abstracts away low-level control complexity so that benchmark performance reflects collaboration capability rather than fine-grained action proficiency. Tasks are designed to be diverse, challenging, and inherently collaborative: they cannot be solved by any single agent and require sustained coordination over multiple rounds.

\subsection{Simulation Environment}

A key challenge in designing a multi-agent collaboration benchmark is choosing the right environment. Existing simulators fall into two extremes. Complex environments like Minecraft offer rich worlds but burden agents with low-level control (precise 3D navigation, block placement, inventory management), making it difficult to isolate collaboration capability from action-control proficiency. Simpler environments like Werewolf or Overcooked provide clean interfaces but lack the world complexity needed for diverse, long-horizon collaboration tasks. Our design targets the middle ground: an environment that is \emph{complex enough} to support diverse collaboration scenarios while \emph{abstracting away} low-level control so that benchmark scores primarily reflect the quality of joint planning and coordination decisions.

\paragraph{Sandbox.} We build upon Kaetram\anontext{\footnote{An open-source MMORPG engine. URL withheld for anonymity.}}{\footnote{\url{https://github.com/Kaetram/Kaetram-Open}}}, an open-source MMORPG engine that provides the world complexity we need: a persistent 2D world spanning $1056 \times 768$ tiles across 9 biome types, with 380+ items, 144 mob types (Level 1 to 250+), 70+ NPCs, and 1,531 harvestable resource nodes across 8 skills (lumberjacking, mining, fishing, foraging, crafting, smithing, fletching, and cooking). This rich content enables diverse task categories (combat, crafting, trading, exploration, etc.) without requiring us to build a game world from scratch.

\paragraph{Agent-Environment Interface.} To ensure that benchmark performance reflects collaboration rather than low-level control, we built a custom abstraction layer on top of Kaetram that exposes 13 high-level API tools (e.g., \texttt{move}, \texttt{attack}, \texttt{harvest}, \texttt{craft}, \texttt{transfer}, \texttt{chat}). Each tool encapsulates complex multi-step game mechanics into a single function call. For instance, \texttt{attack\_entity} handles the entire combat sequence (pathfinding to the target, initiating attack, completing combat, and collecting loot) rather than requiring agents to manage each step individually. Similarly, \texttt{harvest\_resource} handles navigation to the resource node, performing the gathering action, and collecting the result. This design means agents spend their decision budget on \emph{what} to do and \emph{who} to coordinate with, not on \emph{how} to execute low-level actions. Each turn, agents also receive (1)~a structured text observation of nearby tiles, entities, and their own status, and (2)~a task-specific guideline document describing relevant crafting recipes, monster attributes, and resource locations.

\paragraph{Agent Orchestration.} Although Kaetram runs in real time, we convert it into a \emph{turn-based} environment for reproducible evaluation. Each round, every agent sequentially receives a fresh observation, selects one API tool call, and waits for the action to resolve before the next agent acts. This round-robin protocol ensures deterministic turn order. Under the blackbox setting used in all our experiments, agents have no access to other agents' internal states, observations, or action histories. Coordination relies solely on explicit chat messages routed through the environment.

\subsection{Task Design}

Each task in AgentWorld is structured as follows (see Figure~\ref{fig:task_design} for an example):
\begin{itemize}[nosep,leftmargin=1.2em]
    \item A \textbf{primary objective} requiring multi-agent collaboration (e.g., ``craft a magic staff for the wizard''),
    \item \textbf{Agent configurations} with asymmetric roles, skills, spawn locations, and starting items,
    \item A \textbf{round budget} limiting the number of interaction rounds,
    \item \textbf{Task-specific context} where each agent receives a curated document of relevant game mechanics, crafting recipes, and resource locations,
    \item \textbf{A Python-based success judgment function} for determining the success given a trajectory.  
\end{itemize}

Tasks are designed so that no single agent possesses all the skills or resources needed to succeed alone. For example, Task~67 (Resource Caravan) requires 8 agents across three biomes: a leader coordinates two lumberjack teams in the forest, miners in the mountains, and a crafter who receives materials from both groups to forge a pickaxe. Agents must negotiate who gathers what, communicate resource counts across regions, and execute transfers at shared meeting points. Other tasks range from 3-agent crafting chains (Task~1), and 10-agent festival preparation with parallel cooking and smithing (Task~80) to 16-agent continental surveys spanning the entire map (Task~100).

\begin{figure*}[t]
\centering
\includegraphics[width=\textwidth]{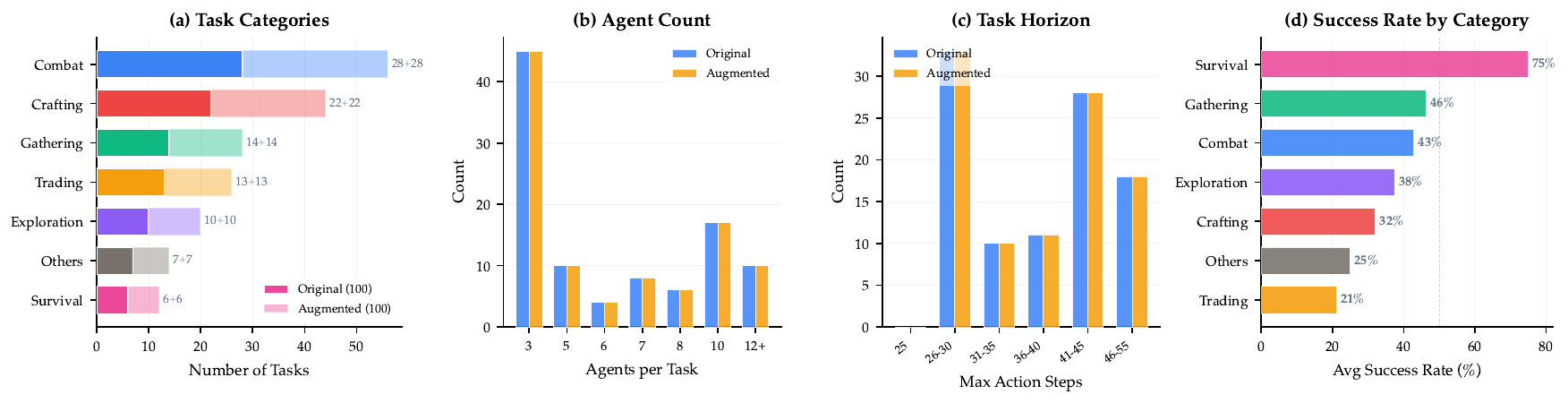}
\caption{\textbf{Dataset Statistics.} (a) Task categories for main and augmented splits. (b) Agents per task. (c) Task horizon (max rounds). (d) Average success rate by category across four models (Gemini 3 Flash, Claude Haiku 4.5, GPT-5 Mini, DeepSeek R1-70B) on the main set.}
\label{fig:dataset_stats}
\end{figure*}

\subsection{Task Annotation}

Five human annotators design tasks following a pre-defined category distribution. For each task, annotators define agent configurations with complementary roles, specify a primary objective with intermediate checkpoints, write a Python verifier that programmatically checks success against the final game state, and curate task-specific game documentation for each agent. Tasks must require genuine multi-agent collaboration (no single agent can succeed alone) and cover diverse collaboration patterns. Each task is verified through pilot experiments with two LLMs; failures are classified as system, task, or agent errors, and tasks are revised until only agent errors remain. The round budget for each task (ranging from 25 to 55) is calibrated from these pilot runs: annotators set an initial estimate, observe how many rounds successful completions require, and adjust the budget to be challenging but feasible. We additionally generate 100 augmented variants using Claude Opus with variations in objectives, spawn locations, and initial items. See Appendix for the full annotation protocol.

\subsection{Dataset Statistics}

The dataset and implementation are publicly available in our \href{https://github.com/openagents-org/agentworld}{GitHub repository}.

The dataset contains 100 human-annotated main tasks and 100 LLM-augmented variants.\footnote{Augmented tasks are generated by Claude Opus 4.5, prompted to follow the structure of human-crafted tasks but with variations in objectives, agent spawn locations, and initial items. Task quality is validated by humans. More discussion is available in \S\ref{app:augmented_task}.} Figure~\ref{fig:dataset_stats} summarizes the statistics. The main tasks span 8 categories (Figure~\ref{fig:dataset_stats}a), with combat and crafting being the most common, and coordination and construction the rarest. Tasks involve 3--20 agents (Figure~\ref{fig:dataset_stats}b), with the majority requiring 3 agents (45 tasks) and a long tail extending to 20-agent tasks. Task horizons range from 25 to 55 maximum allowed rounds (Figure~\ref{fig:dataset_stats}c), with an average budget of 38 rounds. The 100 LLM-augmented variants follow a similar category and horizon distribution to the main tasks (Figure~\ref{fig:dataset_stats}a--c). Average pass rates across all evaluated models (Figure~\ref{fig:dataset_stats}d) reveal that survival tasks are the easiest (67--100\% SR) while coordination (12\%) and construction (20\%) are the most challenging, reflecting the difficulty of tight multi-agent synchronization.

\begin{figure*}[t]
\centering
\includegraphics[width=\textwidth]{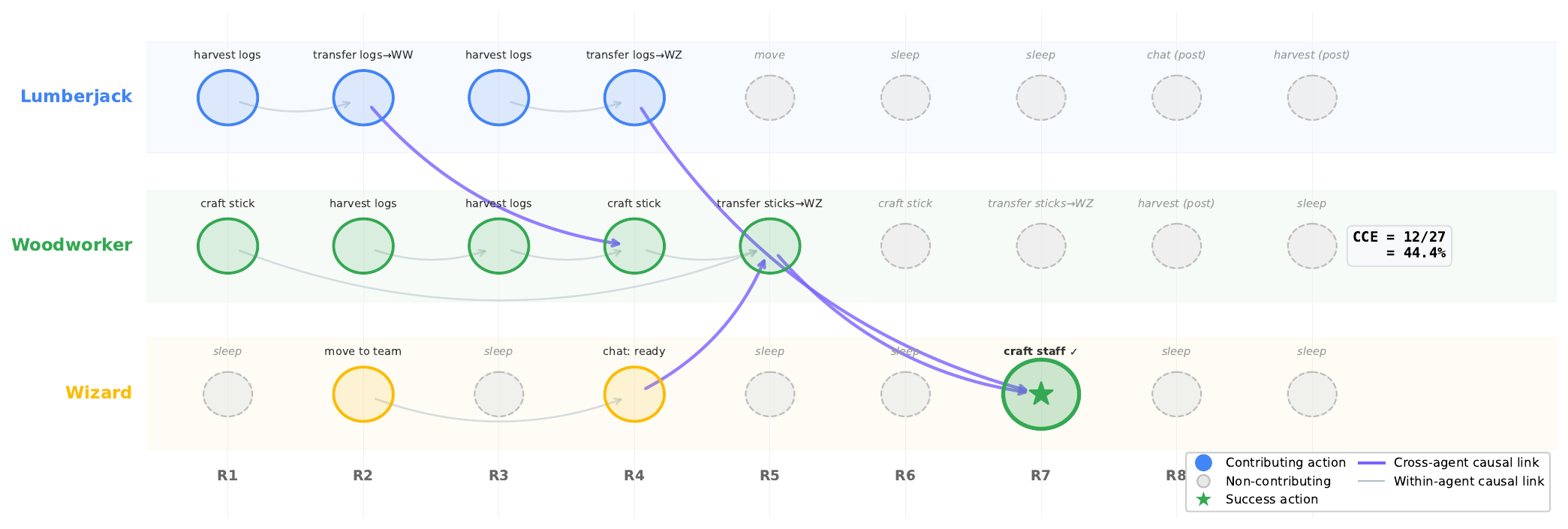}
\vspace{-1em}
\caption{\textbf{CCE illustrated on Task~01.} Colored nodes: contributing actions; gray dashed: non-contributing. Purple arrows: cross-agent causal links. The Lumberjack transfers logs to the Woodworker (R2) and Wizard (R4), enabling downstream crafting. CCE = 12/27 = 44.4\%.}
\label{fig:cce_graph}
\end{figure*}

\subsection{Evaluation Metrics}

Conventionally, the success of multi-agent systems is often evaluated with task success rate (SR), and in the case of AgentWorld, each task couples with its own Python-based function for judging the success. However, as achieving the task success does not always entail a successful collaboration, we propose a new quantitative metric (i.e., causal collaboration effectiveness) for evaluating the quality of cross-agent collaboration.

\paragraph{Task Success Rate (SR)} Based on the primary objective, the success of a task is determined by a Python-based verifier script written for each task. The verifier programmatically checks the final game state (e.g., counting items in inventories, checking kill counts, verifying agent survival). Optionally, the verifier can also check the trajectory if determining the success requires information further than the final game state. A task is successful if and only if all success criteria are met at the end of the trajectory. 

For each task, we also report partial success rate (PSR) as a soft metric using individual checkpoints reported by the verifiers (e.g., ``Logs: 3/5, Kills: 0/3, All alive: True''). PSR is the fraction of checkpoint items achieved, averaged across all tasks. Successful tasks receive PSR = 100\%.

\paragraph{Causal Collaboration Effectiveness (CCE)} SR and PSR measure \emph{what} agents achieved but not \emph{how collaboratively} they achieved it. A task can succeed with high SR yet poor collaboration if a single agent does all the work while others idle. Some existing works adopted LLM-as-judge approaches that score collaboration quality holistically (e.g., ``rate coordination 1--5''), but such judgments are subjective and shift when the judge model is updated, making results difficult to reproduce.

We propose CCE, a graph-based metric that quantifies collaboration quality by constructing a \emph{Causal Action Graph} over task trajectories as illustrated in Figure~\ref{fig:cce_graph}. The algorithm works by iterative backward tracing: we first identify the action(s) that directly achieved the task objective (the success actions), then sweep backward round by round, querying an LLM at each round to determine which actions causally enabled any already-identified contributing action. This BFS-like process accumulates a \emph{contributing set} $\mathcal{C}$ of all actions on a causal path to success. CCE is then the ratio of contributing actions to total actions: $\text{CCE} = |\mathcal{C}|/|\mathcal{T}|$, where $\mathcal{T}$ is the set of all actions taken by all agents across all rounds. For failed tasks, CCE $= 0$ by definition. Crucially, the LLM judge makes only relatively objective binary causal judgments (``did fishing for shrimp enable the later transfer of shrimp?''), not subjective quality assessments (``rate the collaboration 1--5''), making CCE substantially more reproducible across judge model versions. See Appendix~\ref{app:cce} for the full formalization.

\section{Experiments}
\label{sec:experiments}

\subsection{Experimental Setup}

We evaluate four frontier LLMs on the AgentWorld benchmark: \textbf{Gemini 3 Flash}, \textbf{Claude Haiku 4.5}, \textbf{GPT-5 Mini}, and \textbf{DeepSeek R1-70B}. All models use identical system prompts, user prompt templates, and API tool definitions, ensuring a fair comparison of collaborative reasoning capabilities rather than prompt engineering.

We evaluate on the 100 human-annotated main tasks and 100 augmented variants under the blackbox paradigm described in \S\ref{sec:benchmark}. Each model runs all tasks with the same round-based protocol: agents observe, act (one API tool call per turn), and communicate via chat. We report SR, PSR, and CCE as defined in \S\ref{sec:benchmark}. The full agent prompt is provided in Appendix~\ref{app:agent_prompt}.

\subsection{Main Results}

\begin{table*}[t]
\centering
\begin{tabular}{l|rrr|rrr|rr}
\toprule
 & \multicolumn{3}{c|}{\textbf{Main Set}} & \multicolumn{3}{c|}{\textbf{Augmented Set}} & \multicolumn{2}{c}{\textbf{Behavior}} \\
\textbf{Model} & \textbf{SR\%} & \textbf{PSR\%} & \textbf{CCE} & \textbf{SR\%} & \textbf{PSR\%} & \textbf{CCE} & \textbf{Rounds} & \textbf{Chats} \\
\midrule

DeepSeek R1-70B   & 20.0 & 43.5 & 0.125 & 10.0 & 28.9 & 0.063 & 34.9 &  7.5 \\
Claude Haiku 4.5  & 45.0 & 62.7 & 0.294 & 26.0 & 49.1 & \textbf{0.179} & 35.9 & 26.6 \\
GPT-5 Mini        & 36.0 & 62.6 & 0.206 & 21.0 & 41.6 & 0.145 & 34.5 & 44.1 \\
Gemini 3 Flash    & \textbf{52.0} & \textbf{71.5} & \textbf{0.320} & 24.0 & \textbf{61.3} & 0.113 & 26.4 & 11.0 \\
Muse-Glimmer-30B & 38.0 & 45.3 & 0.128 & \textbf{36.0} & 52.5 & 0.126 & 41.65 & 13.58 \\
Gemma 4 26B A4B$^{\dagger}$ & 34.7 & 43.7 & 0.110 & 30.6 & 43.8 & 0.087 & 35.40 & 47.51 \\
\bottomrule
\end{tabular}
\caption{\textbf{Main results on the AgentWorld benchmark.} SR: task success rate (\%). PSR: partial success rate (\%). CCE: causal collaboration effectiveness averaged over all tasks (failed tasks contribute CCE $= 0$). Rounds and Chats are averages on the main set. Bold values identify the best reported scores excluding provisional results. The additional models use the 100-task augmented v1 split. Muse-Glimmer-30B results are final; $^{\dagger}$Gemma results are provisional and averaged over scored tasks.}
\label{tab:model_comparison}
\end{table*}

Table~\ref{tab:model_comparison} presents the main results and two additional evaluations. The following comparisons refer to the four primary models. On the main set, Gemini 3 Flash leads with 52.0\% SR, followed by Claude Haiku 4.5 (45.0\%), GPT-5 Mini (36.0\%), and DeepSeek R1-70B (20.0\%). PSR is consistently higher than SR across all models, ranging from 43.5\% (DeepSeek) to 71.5\% (Gemini). CCE ranges from 0.125 (DeepSeek) to 0.320 (Gemini), indicating that less than a third of all agent actions causally contribute to success even for the best model. Models differ markedly in behavior: Gemini uses the fewest rounds (26.4), DeepSeek uses the fewest chats (7.5), and GPT-5 Mini takes the most actions (131.3) and sends the most messages (44.1). On the augmented set, all models show substantial drops, with DeepSeek falling to just 10.0\% SR. We highlight several findings below.


\paragraph{More communication does not mean better collaboration.} Among the four primary models, GPT-5 Mini sends the most chat messages per task (44.1) but ranks third in SR (36.0\%). DeepSeek R1-70B sends the fewest (7.5) and ranks last (20.0\%). Gemini 3 Flash communicates moderately (11.0) yet achieves the highest SR. On failed tasks, GPT-5 Mini devotes 26\% of its actions to chatting, suggesting it over-communicates at the expense of acting. DeepSeek delays its first chat message until round 5 on average (vs.\ round 1--2 for all other models), likely missing the critical early-coordination window. These patterns suggest an optimal communication regime: enough to coordinate but not so much that it displaces productive action.

\paragraph{Task difficulty varies sharply by category and augmentation.} Across the four primary models on the main set, survival tasks are the easiest (50--100\% SR across models) while coordination (0--50\%), construction (0--60\%), and trading (8--46\%) are the hardest, reflecting the difficulty of tight multi-agent synchronization and negotiation. On the augmented set, all models show substantial SR drops (Gemini: 52$\rightarrow$24\%, Claude: 45$\rightarrow$26\%, GPT-5: 36$\rightarrow$21\%, DeepSeek: 20$\rightarrow$10\%), confirming that augmented tasks are genuinely harder. Augmented tasks also require more rounds and actions, with GPT-5 Mini averaging 152 actions per task vs.\ 131 on the main set.

\subsection{Causal Collaboration Effectiveness}

We apply the CCE metric defined in \S\ref{sec:benchmark} to all trajectories (Table~\ref{tab:model_comparison}, Figure~\ref{fig:cce_graph}).

\paragraph{The vast majority of agent actions do not contribute to success.} Even the best model (Gemini, CCE = 0.320) has less than a third of its total actions on a causal path to task completion. For DeepSeek (CCE = 0.125), nearly 88\% of all actions are wasted. This reveals a fundamental inefficiency: agents spend most of their effort on actions that neither directly achieve objectives nor enable other agents' progress.

\paragraph{CCE tracks SR but adds nuance.} The CCE ranking (Gemini 0.320 $>$ Claude 0.294 $>$ GPT-5 0.206 $>$ DeepSeek 0.125) broadly follows SR, since failed tasks contribute CCE = 0. However, conditioning on successful tasks reveals a different picture: Claude achieves the highest per-task collaboration efficiency (0.653 on successes vs.\ Gemini's 0.609), meaning Claude wastes fewer actions when it does succeed. GPT-5 Mini has the lowest efficiency even on successful tasks (0.571), consistent with its high action count and excessive chatting.

\paragraph{CCE agrees with human causal judgments.} A natural concern is that CCE still relies on an LLM to make the underlying causal judgments. To test whether these judgments are well-posed, we compare the judge against a human annotator. We sample 84 action--contribution judgments drawn from the trajectories of two agent models (Gemini 3 Flash and Claude Haiku 4.5) across 11 tasks. A human annotator independently labels each action as contributing or not, following the same policy given to the judge. The annotator agrees with the GPT-4.1 judge on 69 of 84 judgments (82\% raw agreement), corresponding to a Cohen's $\kappa$ of 0.64 (``substantial'' on the Landis--Koch scale). For reference, two independent judge models agree at a comparable level: recomputing CCE with Claude Sonnet 4 instead of GPT-4.1 yields action-level agreement of $\kappa = 0.66$ (84\% raw) over 7{,}298 decisions. In other words, humans agree with the judge about as strongly as two different judges agree with each other, indicating that the binary causal judgment is well-posed rather than an artifact of a particular model. We adopt an intentionally inclusive labeling policy (\S\ref{app:cce}): marginally helpful actions are counted as contributing, so CCE is best read as an \emph{upper bound} on the fraction of genuinely useful actions.

\paragraph{CCE rankings are robust to the choice of judge model.} A key advantage of CCE over direct LLM-as-judge scoring is that the LLM makes binary causal judgments rather than subjective quality assessments, which is more stable across judge models. To verify this, we recompute CCE with three judges from different families (GPT-4.1, our primary judge; Claude Sonnet 4; and Llama-4 Maverick) and report the results in Table~\ref{tab:cce_judges} (Appendix~\ref{app:extra_results}). The relative ordering of agent models is \emph{fully preserved} under all three judges (Gemini $>$ Claude $>$ GPT-5 $>$ DeepSeek). Absolute CCE values shift by at most $\sim$0.05 (e.g., Claude Haiku moves from 0.294 to 0.241 between GPT-4.1 and Claude Sonnet 4), which does not affect any pairwise comparison. This stability supports our claim that reducing the judge's role to binary causal decisions yields a more reproducible metric than holistic collaboration scoring.

\subsection{Baselines, Ablations, and Robustness}
\label{sec:baselines}

\begin{table}[t!]
\centering
\small
\begin{tabular}{lccc}
\toprule
\textbf{Setting} & \textbf{SR\%} & \textbf{Avg.\ Rounds} & \textbf{Avg.\ Chats} \\
\midrule
Random actions            & 5.7  & 34.1 & 8.3 \\
Shared-plan-no-communication          & 17.6 & 30.7 & 0.0 \\
No-communication          & 22.9 & 32.0 & 0.0 \\
Single-agent              & 28.6 & 28.0 & 0.0 \\
Vanilla multi-agent (LLM) & 52.0 & 26.4 & 11.0 \\
Oracle communication      & 60.0 & 25.2 & 7.9 \\
\bottomrule
\end{tabular}
\caption{\textbf{Baseline comparison} on AgentWorld (Gemini 3 Flash). Random actions use no LLM. Oracle communication grants one agent visibility into all messages to coordinate the team. Vanilla multi-agent is the default blackbox setting reported in Table~\ref{tab:model_comparison}.}
\label{tab:baselines}
\end{table}

\paragraph{Baselines.} To calibrate the difficulty of AgentWorld and disentangle raw LLM capability from multi-agent orchestration, we construct five baselines: (i)~\emph{random actions}, selecting uniformly among legal actions with no LLM; (ii)~\emph{single-agent}, assigning all roles and items to one agent; 
(iii)~\emph{no-communication}, 
where agents act without communicating to other agents by disabling chat and message tools; 
(iv)~\emph{shared-plan-no-communication}, 
which is identical to (iii) except that agents first receive a shared plan written by humans and then execute their assigned part.
and (v)~\emph{oracle communication}, where one agent observes all messages and coordinates the team, establishing an upper bound on communication quality. 


Results are in Table~\ref{tab:baselines}. Random actions solve only 5.7\% of tasks, and the single-agent baseline reaches 28.6\%, confirming that the tasks genuinely require multiple coordinating agents. 
Interestingly, adding a shared plan does not help and slightly hurts (22.9\% vs.\ 17.6\%). Without communication, agents cannot verify teammates' progress or repair the plan when its assumptions break, so committing to a fixed assignment can leave them stalled on steps that depend on others, whereas unplanned agents simply act on what they observe. This suggests that the value of planning in AgentWorld depends on the ability to revise it during execution, which is consistent with the gains from communication discussed below.
Furthermore, oracle communication improves over the vanilla blackbox setting (52.0\%$\rightarrow$60.0\%), showing that better communication helps, yet the 40\% residual failure rate even with oracle communication indicates that much of the difficulty is intrinsic to long-horizon planning and resource allocation, not merely decentralization.

\begin{table}[t]
\centering
\small
\begin{tabular}{lccc}
\toprule
\textbf{Setting} & \textbf{SR\%} & \textbf{Avg.\ Rounds} & \textbf{Avg.\ Chats} \\
\midrule
Full AgentWorld            & 54.3 & 18.8 & 4.7 \\
\;\;w/o asymmetric roles   & 45.7 & 20.2 & 4.5 \\
\;\;w/o long horizon       & 37.1 & 8.7  & 2.4 \\
\;\;w/ random spawn        & 51.4 & 19.5 & 4.6 \\
\;\;w/o documentation      & 29.4 & 30.7 & 3.9 \\
\bottomrule
\end{tabular}
\caption{\textbf{Ablation of key design components} on a 35-task subset (Gemini 3 Flash). Every component contributes; removing task documentation or the long horizon has the largest effect.}
\label{tab:ablations}
\end{table}

\paragraph{Design-component ablations.} We ablate four design choices on a 35-task subset (Table~\ref{tab:ablations}): removing asymmetric role information, shortening the round budget from 50 to 10 rounds, randomizing spawn locations, and withholding task-specific documentation. Every ablation degrades performance, by up to 25 points. The largest drops come from removing documentation (54.3\%$\rightarrow$29.4\%) and shortening the horizon (54.3\%$\rightarrow$37.1\%), confirming that these design elements are load-bearing: without them, agents are bottlenecked by information gathering and time pressure rather than by collaboration itself. Randomizing spawn locations has the smallest effect (54.3\%$\rightarrow$51.4\%). This supports our design goal of isolating collaboration capability from incidental sources of difficulty.

\paragraph{Run-to-run variance.} We run Gemini 3 Flash three times (temperature 0.7; seeds 42, 819, and 314) on the same 35-task subset. Task success rate is 54.3\% in each run (sample standard deviation 0.0 percentage points). Across the three run-level averages, rounds are 18.77, 15.25, and 18.91, while chats are 4.69, 4.50, and 4.34. These yield sample standard deviations of 2.07 rounds and 0.18 chats, respectively, using the $n-1$ denominator. These statistics are calculated from the reported seed-level averages. Identical aggregate SR does not imply identical outcomes on individual tasks; this three-run, single-model analysis does not establish stability across models or statistical significance of model differences.

\subsection{Analysis on Agent-to-Agent Communication}
\label{sec:analysis}

To further understand the communication failures during the collaboration, we inspect the messages generated by Gemini 3 Flash in the traces of 100 tasks and cluster them according to their types. As shown in Table~\ref{tab:agent-failure-modes} (Appendix~\ref{app:extra_results}), the errors of agent communication can be classified into six types. Among them, the most significant one is stale and redundant, covering 37.7\% of the errors. The underlying cause is likely due to incorrect understanding or prediction of other agents' states.

\section{Conclusion}
\label{sec:conclusion}

We presented AgentWorld, a benchmark for evaluating long-horizon multi-agent collaboration. Built on an MMORPG sandbox tailored for agents to play through 13 tools, the benchmark contains 100 human-annotated tasks (with 100 augmented variants) requiring 3--20 agents with asymmetric roles to coordinate through natural language under a blackbox setting where no agent can observe others' internal states. For evaluation, in addition to success rates, we propose Causal Collaboration Effectiveness (CCE), a metric that constructs causal action graphs via backward tracing and measures what fraction of a team's effort actually contributed to the outcome. 

Experiments across four frontier models (Gemini 3 Flash, Claude Haiku 4.5, GPT-5 Mini, DeepSeek R1-70B) reveal several key findings: (1) even the best model achieves only 52.0\% task success, with 27\% of tasks unsolved by any model; (2) CCE analysis shows that less than a third of all agent actions causally contribute to success (CCE = 0.320 for the best model), indicating massive inefficiency in collaboration; (3) a persistent gap between partial and full success (e.g., 71.5\% PSR vs.\ 52.0\% SR for Gemini) shows that agents can start collaborating but fail at late-stage coordination; and (4) communication quantity does not predict success, with the highest-chatting model ranking third and the most efficient model communicating moderately. These results suggest that collaboration remains a common gap in current foundational models, distinct from and harder than single-agent reasoning. 

\section*{Ethics Statement}
AgentWorld is a sandbox environment with simulated scenarios. It is designed to test the collaboration of LLMs rather than negative behaviors such as attacks. The annotation process involves annotators from diverse backgrounds, who are compensated in accordance with local standards.

\section*{Acknowledgments}
The project was supported in part by NSF IIS-2338418, NSF DMS-2533995, and Coefficient Giving (formerly Open Philanthropy). We thank Peakmojo Inc. for supporting the compute resource for this research. We thank volunteers from OpenAgents Community \footnote{\url{https://openagents.org}} for helping on data annotation and evaluation.

\bibliography{mybib}

\begin{thebibliography}{37}
\providecommand{\natexlab}[1]{#1}
\providecommand{\url}[1]{\texttt{#1}}
\expandafter\ifx\csname urlstyle\endcsname\relax
  \providecommand{\doi}[1]{doi: #1}\else
  \providecommand{\doi}{doi: \begingroup \urlstyle{rm}\Url}\fi

\bibitem[Agapiou et~al.(2022)Agapiou, Vezhnevets, Du{\'{e}}nez{-}Guzm{\'{a}}n,
  Matyas, Mao, Sunehag, K{\"{o}}ster, Madhushani, Kopparapu, Comanescu,
  Strouse, Johanson, Singh, Haas, Mordatch, Mobbs, and
  Leibo]{DBLP:journals/corr/abs-2211-13746}
John~P. Agapiou, Alexander~Sasha Vezhnevets, Edgar~A.
  Du{\'{e}}nez{-}Guzm{\'{a}}n, Jayd Matyas, Yiran Mao, Peter Sunehag, Raphael
  K{\"{o}}ster, Udari Madhushani, Kavya Kopparapu, Ramona Comanescu,
  DJ~Strouse, Michael~Bradley Johanson, Sukhdeep Singh, Julia Haas, Igor
  Mordatch, Dean Mobbs, and Joel~Z. Leibo.
\newblock Melting pot 2.0.
\newblock \emph{CoRR}, abs/2211.13746, 2022.
\newblock URL \url{https://arxiv.org/abs/2211.13746}.

\bibitem[Agashe et~al.(2023)Agashe, Fan, Reyna, and
  Wang]{agashe2023llmcoordination}
Saaket Agashe, Yue Fan, Anthony Reyna, and Xin~Eric Wang.
\newblock Evaluating multi-agent coordination abilities in large language
  models.
\newblock \emph{arXiv preprint arXiv:2310.03903}, 2023.

\bibitem[AL et~al.(2024)AL, Ahn, Becker, Carroll, Christie, Cortes, Demirci,
  Du, Li, Luo, Wang, Willows, Yang, and Yang]{altera2024projectsid}
Altera. AL, Andrew Ahn, Nic Becker, Stephanie Carroll, Nico Christie, Manuel
  Cortes, Arda Demirci, Melissa Du, Frankie Li, Shuying Luo, Peter~Y Wang,
  Mathew Willows, Feitong Yang, and Guangyu~Robert Yang.
\newblock Project sid: Many-agent simulations toward ai civilization.
\newblock \emph{arXiv preprint arXiv:2411.00114}, 2024.

\bibitem[Bailis et~al.(2024)Bailis, Friedhoff, and
  Chen]{bailis2024werewolfarena}
Suma Bailis, Jane Friedhoff, and Feiyang Chen.
\newblock Werewolf arena: A case study in llm evaluation via social deduction.
\newblock \emph{arXiv preprint arXiv:2407.13943}, 2024.

\bibitem[Bara et~al.(2021)Bara, Sky, and Chai]{bara2021mindcraft}
Cristian-Paul Bara, CH-Wang Sky, and Joyce Chai.
\newblock Mindcraft: Theory of mind modeling for situated dialogue in
  collaborative tasks.
\newblock In \emph{Proceedings of the 2021 Conference on Empirical Methods in
  Natural Language Processing}, pp.\  1112--1125, 2021.

\bibitem[Bard et~al.(2020)Bard, Foerster, Chandar, Burch, Lanctot, Song,
  Parisotto, Dumoulin, Moitra, Hughes, Dunning, Mourad, Larochelle, Bellemare,
  and Bowling]{DBLP:journals/corr/abs-1902-00506}
Nolan Bard, Jakob~N. Foerster, Sarath Chandar, Neil Burch, Marc Lanctot,
  H.~Francis Song, Emilio Parisotto, Vincent Dumoulin, Subhodeep Moitra, Edward
  Hughes, Iain Dunning, Shibl Mourad, Hugo Larochelle, Marc~G. Bellemare, and
  Michael Bowling.
\newblock The {Hanabi} challenge: {A} new frontier for {AI} research.
\newblock \emph{Artificial Intelligence}, 280:\penalty0 103216, 2020.
\newblock \doi{10.1016/j.artint.2019.103216}.

\bibitem[Carroll et~al.(2019)Carroll, Shah, Ho, Griffiths, Seshia, Abbeel, and
  Dragan]{DBLP:journals/corr/abs-1910-05789}
Micah Carroll, Rohin Shah, Mark~K. Ho, Thomas~L. Griffiths, Sanjit~A. Seshia,
  Pieter Abbeel, and Anca~D. Dragan.
\newblock On the utility of learning about humans for human-{AI} coordination.
\newblock \emph{CoRR}, abs/1910.05789, 2019.
\newblock URL \url{http://arxiv.org/abs/1910.05789}.

\bibitem[Chen et~al.(2023)Chen, Su, Zuo, Yang, Yuan, Qian, Chan, Qin, Lu, Xie,
  Liu, Sun, and Zhou]{chen2023agentverse}
Weize Chen, Yusheng Su, Jingwei Zuo, Cheng Yang, Chenfei Yuan, Chen Qian,
  Chi-Min Chan, Yujia Qin, Yaxi Lu, Ruobing Xie, Zhiyuan Liu, Maosong Sun, and
  Jie Zhou.
\newblock Agentverse: Facilitating multi-agent collaboration and exploring
  emergent behaviors.
\newblock \emph{arXiv preprint arXiv:2308.10848}, 2023.

\bibitem[Costarelli et~al.(2024)Costarelli, Allen, Hauksson, Sodunke,
  Hariharan, Cheng, Li, Clymer, and Yadav]{costarelli2024gamebench}
Anthony Costarelli, Mat Allen, Roman Hauksson, Grace Sodunke, Suhas Hariharan,
  Carlson Cheng, Wenjie Li, Joshua Clymer, and Arjun Yadav.
\newblock Gamebench: Evaluating strategic reasoning abilities of llm agents.
\newblock \emph{arXiv preprint arXiv:2406.06613}, 2024.

\bibitem[Fan et~al.(2022)Fan, Wang, Jiang, Mandlekar, Yang, Zhu, Tang, Huang,
  Zhu, and Anandkumar]{fan2022minedojo}
Linxi Fan, Guanzhi Wang, Yunfan Jiang, Ajay Mandlekar, Yuncong Yang, Haoyi Zhu,
  Andrew Tang, De-An Huang, Yuke Zhu, and Anima Anandkumar.
\newblock Minedojo: Building open-ended embodied agents with internet-scale
  knowledge.
\newblock In \emph{Advances in Neural Information Processing Systems
  (NeurIPS)}, 2022.

\bibitem[Gong et~al.(2024)Gong, Huang, Ma, Noda, Durante, Zheng, Terzopoulos,
  Fei-Fei, Gao, and Vo]{gong2024mindagent}
Ran Gong, Qiuyuan Huang, Xiaojian Ma, Yusuke Noda, Zane Durante, Zilong Zheng,
  Demetri Terzopoulos, Li~Fei-Fei, Jianfeng Gao, and Hoi Vo.
\newblock Mindagent: Emergent gaming interaction.
\newblock In \emph{Findings of the Association for Computational Linguistics:
  NAACL 2024}, pp.\  3154--3183, 2024.

\bibitem[Guo et~al.(2024)Guo, Chen, Wang, Chang, Pei, Chawla, Wiest, and
  Zhang]{guo2024llmmultiagentssurvey}
Taicheng Guo, Xiuying Chen, Yaqi Wang, Ruidi Chang, Shichao Pei, Nitesh~V
  Chawla, Olaf Wiest, and Xiangliang Zhang.
\newblock Large language model based multi-agents: A survey of progress and
  challenges.
\newblock 2024.

\bibitem[Hafner(2021)]{hafner2021crafter}
Danijar Hafner.
\newblock Benchmarking the spectrum of agent capabilities.
\newblock \emph{arXiv preprint arXiv:2109.06780}, 2021.

\bibitem[Leibo et~al.(2021)Leibo, Du{\'{e}}{\~{n}}ez{-}Guzm{\'{a}}n,
  Vezhnevets, Agapiou, Sunehag, Koster, Matyas, Beattie, Mordatch, and
  Graepel]{DBLP:conf/icml/LeiboDVASKMBMG21}
Joel~Z. Leibo, Edgar~A. Du{\'{e}}{\~{n}}ez{-}Guzm{\'{a}}n, Alexander~Sasha
  Vezhnevets, John~P. Agapiou, Peter Sunehag, Raphael Koster, Jayd Matyas,
  Charlie Beattie, Igor Mordatch, and Thore Graepel.
\newblock Scalable evaluation of multi-agent reinforcement learning with
  melting pot.
\newblock In \emph{Proc.\ of the 38th International Conference on Machine
  Learning (ICML)}, volume 139 of \emph{Proceedings of Machine Learning
  Research}, pp.\  6187--6199. PMLR, 2021.

\bibitem[Li et~al.(2023)Li, Hammoud, Itani, Khizbullin, and
  Ghanem]{li2023camel}
Guohao Li, Hasan Abed Al~Kader Hammoud, Hani Itani, Dmitrii Khizbullin, and
  Bernard Ghanem.
\newblock Camel: Communicative agents for "mind" exploration of large language
  model society.
\newblock In \emph{Advances in Neural Information Processing Systems
  (NeurIPS)}, 2023.

\bibitem[Liu et~al.(2023)Liu, Yu, Zhang, Xu, Lei, Lai, Gu, Ding, Men, Yang,
  Zhang, Deng, Zeng, Du, Zhang, Shen, Zhang, Su, Sun, Huang, Dong, and
  Tang]{liu2023agentbench}
Xiao Liu, Hao Yu, Hanchen Zhang, Yifan Xu, Xuanyu Lei, Hanyu Lai, Yu~Gu,
  Hangliang Ding, Kaiwen Men, Kejuan Yang, Shudan Zhang, Xiang Deng, Aohan
  Zeng, Zhengxiao Du, Chenhui Zhang, Sheng Shen, Tianjun Zhang, Yu~Su, Huan
  Sun, Minlie Huang, Yuxiao Dong, and Jie Tang.
\newblock Agentbench: Evaluating llms as agents.
\newblock \emph{arXiv preprint arXiv:2308.03688}, 2023.

\bibitem[Long et~al.(2024)Long, Li, Gong, Wu, Terzopoulos, and
  Gao]{long2024teamcraft}
Qian Long, Zhi Li, Ran Gong, Ying~Nian Wu, Demetri Terzopoulos, and Xiaofeng
  Gao.
\newblock Teamcraft: A benchmark for multi-modal multi-agent systems in
  minecraft.
\newblock \emph{arXiv preprint arXiv:2412.05255}, 2024.

\bibitem[Ma et~al.(2024)Ma, Zhang, Zhu, Yang, Yang, Jin, Lan, Kong, and
  He]{ma2024agentboard}
Chang Ma, Junlei Zhang, Zhihao Zhu, Cheng Yang, Yujiu Yang, Yaohui Jin,
  Zhenzhong Lan, Lingpeng Kong, and Junxian He.
\newblock Agentboard: An analytical evaluation board of multi-turn llm agents.
\newblock In \emph{Advances in Neural Information Processing Systems
  (NeurIPS)}, 2024.

\bibitem[Mandi et~al.(2023)Mandi, Jain, and Song]{mandi2023roco}
Zhao Mandi, Shreeya Jain, and Shuran Song.
\newblock Roco: Dialectic multi-robot collaboration with large language models.
\newblock \emph{arXiv preprint arXiv:2307.04738}, 2023.

\bibitem[Matthews et~al.(2024)Matthews, Beukman, Ellis, Samvelyan, Jackson,
  Coward, and Foerster]{matthews2024craftax}
Michael Matthews, Michael Beukman, Benjamin Ellis, Mikayel Samvelyan, Matthew
  Jackson, Samuel Coward, and Jakob Foerster.
\newblock Craftax: A lightning-fast benchmark for open-ended reinforcement
  learning.
\newblock In \emph{Proceedings of the 41st International Conference on Machine
  Learning (ICML)}, 2024.

\bibitem[Mohammadi et~al.(2025)Mohammadi, Li, Lo, and
  Yip]{mohammadi2025llmagentevalsurvey}
Mahmoud Mohammadi, Yipeng Li, Jane Lo, and Wendy Yip.
\newblock Evaluation and benchmarking of llm agents: A survey.
\newblock 2025.

\bibitem[Paglieri et~al.(2024)Paglieri, Cupial, Coward, Piterbarg, Wolczyk,
  Khan, Pignatelli, Kuci{\'n}ski, Pinto, Fergus, Foerster, Parker-Holder, and
  Rockt{\"a}schel]{paglieri2024balrog}
Davide Paglieri, Bartlomiej Cupial, Samuel Coward, Ulyana Piterbarg, Maciej
  Wolczyk, Akbir Khan, Eduardo Pignatelli, {\L}ukasz Kuci{\'n}ski, Lerrel
  Pinto, Rob Fergus, Jakob~N Foerster, Jack Parker-Holder, and Tim
  Rockt{\"a}schel.
\newblock Balrog: Benchmarking agentic llm and vlm reasoning on games.
\newblock \emph{arXiv preprint arXiv:2411.13543}, 2024.

\bibitem[Park et~al.(2023)Park, O'Brien, Cai, Morris, Liang, and
  Bernstein]{park2023generative}
Joon~Sung Park, Joseph~C O'Brien, Carrie~J Cai, Meredith~Ringel Morris, Percy
  Liang, and Michael~S Bernstein.
\newblock Generative agents: Interactive simulacra of human behavior.
\newblock \emph{Proceedings of the 36th annual ACM symposium on user interface
  software and technology}, pp.\  1--22, 2023.

\bibitem[Park et~al.(2024)Park, Zou, Shaw, Hill, Cai, Morris, Liang, and
  Bernstein]{park2024generativeagents1000}
Joon~Sung Park, Carolyn Zou, Aaron Shaw, Benjamin Hill, Carrie Cai,
  Meredith~Ringel Morris, Percy Liang, and Michael~S Bernstein.
\newblock Generative agent simulations of 1,000 people.
\newblock \emph{arXiv preprint arXiv:2411.10109}, 2024.

\bibitem[Piao et~al.(2025)Piao, Yan, Zhang, Li, Yan, Lan, Lu, Zheng, Wang,
  Zhou, Gao, Xu, Zhang, Rong, Su, and Li]{piao2025agentsociety}
Jinghua Piao, Yuwei Yan, Jun Zhang, Nian Li, Junbo Yan, Xiaochong Lan, Zhihong
  Lu, Zhiheng Zheng, Jing~Yi Wang, Di~Zhou, Chen Gao, Fengli Xu, Fang Zhang,
  Ke~Rong, Jun Su, and Yong Li.
\newblock Agentsociety: Large-scale simulation of llm-driven generative agents
  advances understanding of human behaviors and society.
\newblock \emph{arXiv preprint arXiv:2502.08691}, 2025.

\bibitem[Riedl(2025)]{riedl2025emergentcoordination}
Christoph Riedl.
\newblock Emergent coordination in multi-agent language models.
\newblock \emph{arXiv preprint arXiv:2510.05174}, 2025.

\bibitem[Samvelyan et~al.(2019)Samvelyan, Rashid, de~Witt, Farquhar, Nardelli,
  Rudner, Hung, Torr, Foerster, and Whiteson]{samvelyan19smac}
Mikayel Samvelyan, Tabish Rashid, Christian~Schroeder de~Witt, Gregory
  Farquhar, Nantas Nardelli, Tim G.~J. Rudner, Chia{-}Man Hung, Philip H.~S.
  Torr, Jakob Foerster, and Shimon Whiteson.
\newblock The starcraft multi-agent challenge.
\newblock \emph{CoRR}, abs/1902.04043, 2019.
\newblock URL \url{http://arxiv.org/abs/1902.04043}.

\bibitem[Sun et~al.(2025)Sun, Zhang, Niu, Ren, Xu, Fu, Zhao, Yuan, and
  Wang]{sun2025collab}
Haochen Sun, Shuwen Zhang, Lujie Niu, Lei Ren, Hao Xu, Hao Fu, Fangkun Zhao,
  Caixia Yuan, and Xiaojie Wang.
\newblock Collab-overcooked: Benchmarking and evaluating large language models
  as collaborative agents.
\newblock In \emph{Proceedings of the 2025 Conference on Empirical Methods in
  Natural Language Processing}, pp.\  4922--4951, 2025.

\bibitem[Terry et~al.(2020)Terry, Black, Grammel, Jayakumar, Hari, Sullivan,
  Santos, Perez, Horsch, Dieffendahl, Williams, Lokesh, and
  Ravi]{DBLP:journals/corr/abs-2009-14471}
J.~K. Terry, Benjamin Black, Nathaniel Grammel, Mario Jayakumar, Ananth Hari,
  Ryan Sullivan, Luis Santos, Rodrigo Perez, Caroline Horsch, Clemens
  Dieffendahl, Niall~L. Williams, Yashas Lokesh, and Praveen Ravi.
\newblock Pettingzoo: Gym for multi-agent reinforcement learning.
\newblock \emph{CoRR}, abs/2009.14471, 2020.
\newblock URL \url{https://arxiv.org/abs/2009.14471}.

\bibitem[Tran et~al.(2025)Tran, Dao, Nguyen, Pham, O'Sullivan, and
  Nguyen]{zhu2025multiagentcollabmechanisms}
Khanh-Tung Tran, Dung Dao, Minh-Duong Nguyen, Quoc-Viet Pham, Barry O'Sullivan,
  and Hoang~D. Nguyen.
\newblock Multi-agent collaboration mechanisms: A survey of llms.
\newblock \emph{arXiv preprint arXiv:2501.06322}, 2025.

\bibitem[Wang et~al.(2023)Wang, Xie, Jiang, Mandlekar, Xiao, Zhu, Fan, and
  Anandkumar]{wang2023voyager}
Guanzhi Wang, Yuqi Xie, Yunfan Jiang, Ajay Mandlekar, Chaowei Xiao, Yuke Zhu,
  Linxi Fan, and Anima Anandkumar.
\newblock Voyager: An open-ended embodied agent with large language models.
\newblock \emph{arXiv preprint arXiv:2305.16291}, 2023.

\bibitem[Wu et~al.(2024)Wu, Bansal, Zhang, Wu, Li, Zhu, Jiang, Zhang, Zhang,
  Liu, Awadallah, White, Burger, and Wang]{wu2024autogen}
Qingyun Wu, Gagan Bansal, Jieyu Zhang, Yiran Wu, Beibin Li, Erkang Zhu,
  Li~Jiang, Xiaoyun Zhang, Shaokun Zhang, Jiale Liu, Ahmed~Hassan Awadallah,
  Ryen~W White, Doug Burger, and Chi Wang.
\newblock Autogen: Enabling next-gen llm applications via multi-agent
  conversation.
\newblock In \emph{Conference on Language Modeling (COLM)}, 2024.

\bibitem[Xu et~al.(2024{\natexlab{a}})Xu, Song, Li, Tang, Jain, Bao, Wang,
  Zhou, Guo, Cao, Yang, Lu, Martin, Su, Maben, Mehta, Chi, Jang, Xie, Zhou, and
  Neubig]{xu2024theagentcompany}
Frank~F. Xu, Yufan Song, Boxuan Li, Yuxuan Tang, Kritanjali Jain, Mengxue Bao,
  Zora~Z. Wang, Xuhui Zhou, Zhitong Guo, Murong Cao, Mingyang Yang, Hao~Yang
  Lu, Amaad Martin, Zhe Su, Leander Maben, Raj Mehta, Wayne Chi, Lawrence Jang,
  Yiqing Xie, Shuyan Zhou, and Graham Neubig.
\newblock Theagentcompany: Benchmarking llm agents on consequential real world
  tasks.
\newblock \emph{arXiv preprint arXiv:2412.14161}, 2024{\natexlab{a}}.

\bibitem[Xu et~al.(2024{\natexlab{b}})Xu, Yu, Fang, Wang, and
  Wu]{xu2024werewolflanguageagents}
Zelai Xu, Chao Yu, Fei Fang, Yu~Wang, and Yi~Wu.
\newblock Language agents with reinforcement learning for strategic play in the
  werewolf game.
\newblock In \emph{Proceedings of the 41st International Conference on Machine
  Learning (ICML)}, 2024{\natexlab{b}}.

\bibitem[Yehudai et~al.(2025)Yehudai, Eden, Li, Uziel, Zhao, Bar-Haim, Cohan,
  and Shmueli-Scheuer]{zhao2025llmagentsurvey}
Asaf Yehudai, Lilach Eden, Alan Li, Guy Uziel, Yilun Zhao, Roy Bar-Haim, Arman
  Cohan, and Michal Shmueli-Scheuer.
\newblock Survey on evaluation of llm-based agents.
\newblock \emph{arXiv preprint arXiv:2503.16416}, 2025.

\bibitem[Yu et~al.(2024)Yu, Fu, Deng, and Han]{yu2024mineland}
Xianhao Yu, Jiaqi Fu, Renjia Deng, and Wenjuan Han.
\newblock Mineland: Simulating large-scale multi-agent interactions with
  limited multimodal senses and physical needs.
\newblock \emph{arXiv preprint arXiv:2403.19267}, 2024.

\bibitem[Zhu et~al.(2025)Zhu, Du, Hong, Yang, Guo, Wang, Wang, Qian, Tang, Ji,
  and You]{zhu2025multiagentbench}
Kunlun Zhu, Hongyi Du, Zhaochen Hong, Xiaocheng Yang, Shuyi Guo, Zhe Wang,
  Zhenhailong Wang, Cheng Qian, Xiangru Tang, Heng Ji, and Jiaxuan You.
\newblock Multiagentbench: Evaluating the collaboration and competition of llm
  agents.
\newblock In \emph{Proceedings of the 63rd Annual Meeting of the Association
  for Computational Linguistics (Volume 1: Long Papers)}, pp.\  8580--8622,
  2025.

\end{thebibliography}
\bibliographystyle{colm2026_conference}

\appendix

\section{Additional Experimental Results}
\label{app:extra_results}

Table~\ref{tab:cce_judges} reports CCE under three different judge models (referenced in \S\ref{sec:experiments}), 
and Table~\ref{tab:agent-failure-modes} reports the communication failure-mode analysis (referenced in \S\ref{sec:analysis}).

\begin{table*}[h]
  \centering
  \small
  \begin{tabularx}{\textwidth}{@{} p{2cm} p{4cm} p{4cm} r r @{}}
      \toprule
      \textbf{Category} & \textbf{Definition} & \textbf{Representative Example (Chat Snippet)} & \textbf{Count} & \textbf{Ratio} \\
      \midrule
      Stale \& Redundant & Referencing completed tasks, repeating known states, or asking resolved questions. & Wizard had already crafted the staff in previous round ``... do you have the sticks now?'' & 23 & 37.7\% \\
      \addlinespace
      Misidentification & Incorrect self-role labeling or requesting resources from the wrong agent. & ... coming from the miner, smelter: ``... send 1x coal to me.'' & 10 & 16.4\% \\
      \addlinespace
      Factual Error & Reporting incorrect item names, coordinates, or false readiness of a process. & ``Transferred 12 gold ore (as goldnugget)... smith needs EXACTLY goldore.'' & 9 & 14.8\% \\
      \addlinespace
      Premature Completion & Unilaterally declaring the task finished before critical final steps are verified. & ``All 7 jewelry items are finished... (before ring jeweler's actions shown).'' & 7 & 11.5\% \\
      \addlinespace
      Non-Material Output & Polite but idle chatter (e.g., ``cheering'') without providing coordinates or items. & ``... let me know if you need any help.'' & 7 & 11.5\% \\
      \addlinespace
      Logical Inconsistency & Pursuing invalid strategies or issuing contradictory commands. & Everyone is already dead``... I'm searching for bowls... This is going to be tough alone.'' & 5 & 8.2\% \\
      \midrule
      \textbf{Total} & & & \textbf{61} & \textbf{100\%} \\
      \bottomrule
  \end{tabularx}
  \caption{Failure mode analysis of multi-agent collaboration: a communication perspective.}
  \label{tab:agent-failure-modes}
\end{table*}

\begin{table}[h]
\centering
\small
\begin{tabular}{lccc}
\toprule
\textbf{Agent model} & \textbf{GPT-4.1} & \textbf{Claude Sonnet 4} & \textbf{Llama-4 Maverick} \\
\midrule
Gemini 3 Flash    & 0.320 & 0.338 & 0.330 \\
Claude Haiku 4.5  & 0.294 & 0.241 & 0.284 \\
GPT-5 Mini        & 0.206 & 0.221 & 0.224 \\
DeepSeek R1-70B   & 0.125 & 0.104 & 0.147 \\
\bottomrule
\end{tabular}
\caption{\textbf{CCE under three different judge models} (main set, averaged over all tasks). The ranking of agent models is fully preserved across all three judges; absolute values shift by at most $\sim$0.05.}
\label{tab:cce_judges}
\end{table}

\section{Example Task Definitions}
\label{app:task_examples}

We provide two complete task definitions to illustrate the benchmark format: a crafting task (Task~01, 3 agents) and a combat task (Task~76, 10 agents).

\subsection{Task 01: Cooperative Magic Staff Crafting}

\begin{small}
\begin{verbatim}
task:
  name: "Cooperative Magic Staff Crafting"
objectives:
  primary: "Prepare wizard with crafted magic staff"
max_action_steps: 25

agent_1:
  username: t01_lumberjack_agent
  location: {x: 270, y: 80}
  skill_levels:
    lumberjacking: 25, strength: 25, health: 25
  inventory: [{flask: 10}, {apple: 15}]
  equipment: [axe, leatherarmor, leatherboots]

agent_2:
  username: t01_woodworker_agent
  location: {x: 275, y: 85}
  skill_levels:
    fletching: 15, crafting: 10, strength: 20
  inventory: [{flask: 8}, {apple: 12}]
  equipment: [leatherarmor, leatherboots]

agent_3:
  username: t01_wizard_agent
  location: {x: 280, y: 90}
  skill_levels:
    crafting: 15, magic: 25, health: 25
  inventory: [{flask: 10}, {apple: 15}, {bead: 1}]
  equipment: [leatherarmor, leatherboots]

relevant_game_context: |
  Crafting Recipe - stick:
    craft_item skill=Fletching itemKey=stick
    Materials: 1x logs -> 4x stick
  Crafting Recipe - staff:
    craft_item skill=Crafting itemKey=staff
    Materials: 5x stick + 1x bead -> 1x staff
\end{verbatim}
\end{small}

\paragraph{Verifier Script (Task 01).}
\begin{small}
\begin{verbatim}
def verify(traj_json):
    inventories = get_final_inventories(traj_json)
    staff = has_item_in_any_inventory(
        inventories, 'staff')
    return (1 if staff else 0,
            f"Staff crafted: {staff}")
\end{verbatim}
\end{small}

\subsection{Task 76: Elemental Bosses (10 agents)}

\begin{small}
\begin{verbatim}
task:
  name: "Elemental Bosses - Multi-Boss Combat"
objectives:
  primary: "Defeat the Ice Guardian at (259,151),
    Iron Ogre at (238,48), and
    Water Guardian at (197,633)"
max_action_steps: 48

# 10 agents with combat-focused skills
agent_1:  # Ice team leader
  username: t76_ice_leader_agent
  location: {x: 259, y: 155}
  skill_levels: {strength: 40, defense: 38,
                 health: 50}
  equipment: [heavysword, platearmor, leatherboots]

agent_2:  # Ice team DPS
  username: t76_ice_dps1_agent
  location: {x: 261, y: 153}
  skill_levels: {strength: 42, accuracy: 38,
                 health: 45}
  equipment: [heavysword, platearmor, leatherboots]

agent_3:  # Ice team support
  username: t76_ice_support_agent
  location: {x: 257, y: 153}
  skill_levels: {magic: 35, health: 40, defense: 30}
  equipment: [staff, leatherarmor, leatherboots]

# ... (7 more agents across mountain and water
#  regions with similar structure)
\end{verbatim}
\end{small}

\paragraph{Verifier Script (Task 76).}
\begin{small}
\begin{verbatim}
def verify(traj_json):
    kills = count_combat_kills(traj_json,
        ['Ice Guardian', 'Iron Ogre',
         'Water Guardian'])
    return (1 if kills >= 3 else 0,
            f"Boss kills: {kills}/3")
\end{verbatim}
\end{small}

\section{Task Annotation Platform}
\label{app:annotation_platform}

\begin{figure}[h]
\centering
\includegraphics[width=\columnwidth]{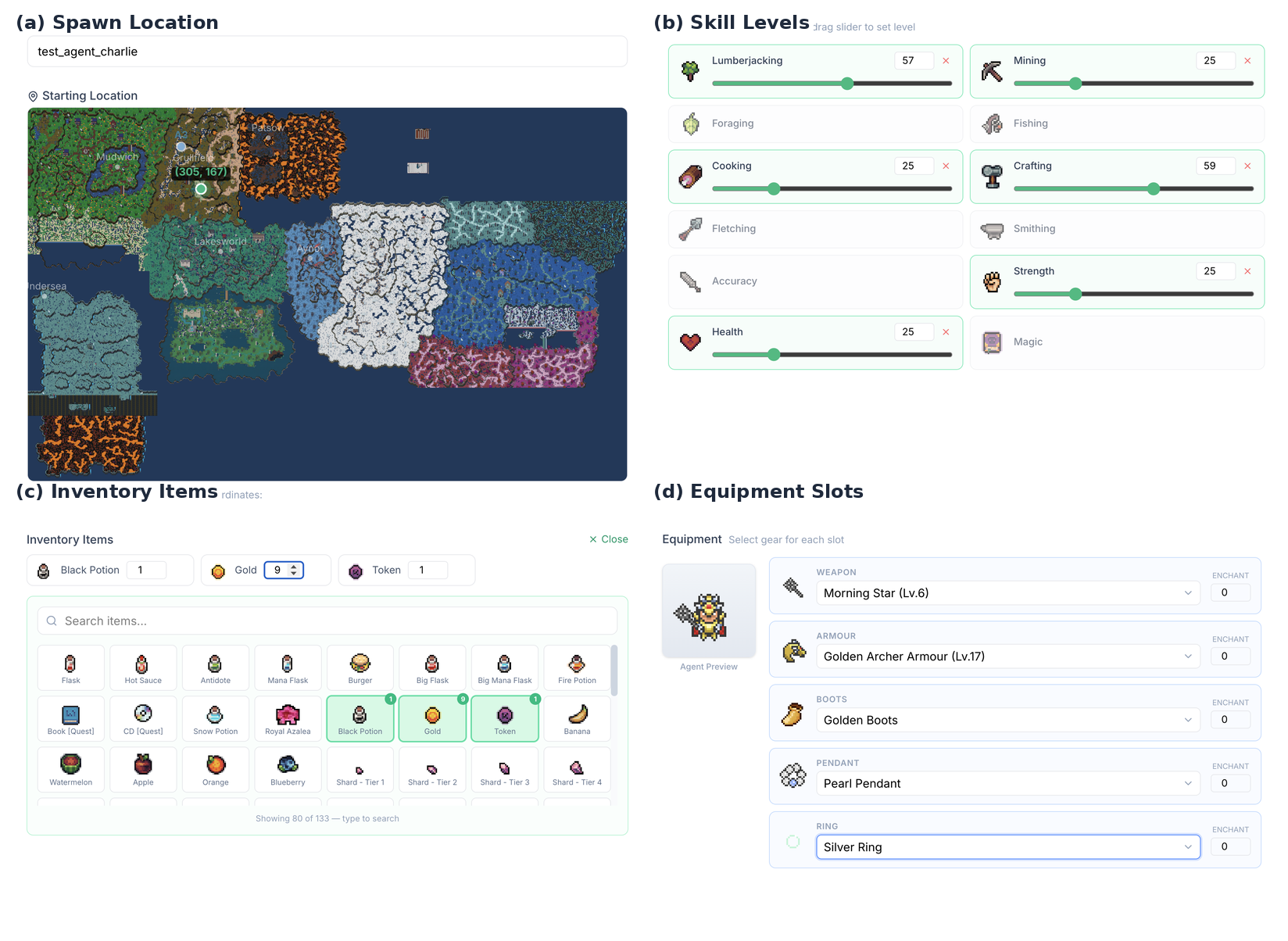}
\caption{\textbf{Task annotation platform.} (a) World map for selecting agent spawn locations. (b) Skill level configuration with sliders. (c) Searchable inventory item browser (133 items). (d) Equipment slots with character preview.}
\label{fig:annotation_platform}
\end{figure}

AgentWorld includes a web-based task annotation platform (Figure~\ref{fig:annotation_platform}) that supports the full annotation workflow described in \S\ref{sec:benchmark}. Annotators configure each agent through an interactive interface that includes:
\begin{itemize}
    \item A world map for selecting agent spawn locations by clicking on specific coordinates,
    \item Skill level sliders with game sprite icons for each of the 8 skills (lumberjacking, mining, foraging, fishing, cooking, crafting, fletching, smithing),
    \item An inventory selector with searchable item browser (133 items available),
    \item An equipment configurator with character preview showing weapon, armor, boots, pendant, and ring slots.
\end{itemize}

\section{Augmented Task}
\label{app:augmented_task}
To analyze the preservation of collaboration for 100 task variants, we further conducted a human validation. We sample 10\% of augmented variants, spanning gathering, construction, combat, and crafting categories. For each task, we ask the annotator to read and compare the original and augmented data files. The results show that 100\% sampled variants preserve the need for multi-agent collaboration. Besides, annotations show 100\% variants preserve the original task objective at the semantic level, and 100\% preserve role asymmetry among agents. The difficulty is also appropriate for 90\% variants, with the remaining cases marked as dependent on whether the substituted enemy was actually weaker than the original one.

Furthermore, our design of task variant creation intentionally avoids breaking the collaboration requirements. For instance, the variant includes heavier objectives (craft more items), more distant spawn locations, and fewer initial items (more difficult to attack mob). These changes do not influence the original execution order of the tasks, thus preserving the original collaboration properties.

\section{Agent Prompt}
\label{app:agent_prompt}

All four models use identical system and user prompts. Below we reproduce the full prompts used in our experiments.

\subsection{System Prompt}

\begin{small}
\begin{verbatim}
You are an intelligent AI agent that plays the
AgentWorld MMORPG game. Your goal is to explore,
interact, collect resources, and engage with the
game world intelligently.

CRITICAL RESPONSE RULE: You MUST call exactly ONE
tool function in every response. Never respond
without calling a tool function.

ACTION-ORIENTED MINDSET:
- When you need to find something specific, use
  tools to actively search or move to areas where
  you might find them
- Be decisive and take action rather than just
  describing what you plan to do

You have access to various game tools through
function calling. Use these tools strategically to:
1. Move around to explore the game world
2. Collect resources when available
3. Interact with other players through chat
4. Engage in combat when appropriate
5. Equip items to improve your character
6. Use 'sleep' only when waiting serves a purpose

COMBAT GUIDELINES:
- The attack_entity function handles ALL aspects of
  combat automatically: movement, attack, and loot
  collection
- Simply use attack_entity with the target's
  instance ID

=== CURRENT ENVIRONMENT OBSERVATION ===
{observation}
=== END OBSERVATION ===

=== CHAT HISTORY ===
{chat_messages}
=== END CHAT HISTORY ===
\end{verbatim}
\end{small}

\subsection{User Prompt (Per-Task, Per-Agent)}

\begin{small}
\begin{verbatim}
**Task:** {{task_name}}

**Description:** {{task_description}}

**Primary Objective:** {{primary_objective}}

**Secondary Objectives:**
{% for obj in secondary_objectives %}
- {{obj}}
{% endfor %}

**Relevant Context:**
{{relevant_game_context}}

**Team Information:**
- Total agents in party: {{total_agents}}
- Your username: {{agent_username}}
- Other team members: {{other_agent_usernames}}
\end{verbatim}
\end{small}

The user prompt is rendered using Jinja2 templates with task-specific variables from the YAML task definition. Each agent receives its own username, the list of teammates, and the task-specific game context (crafting recipes, monster attributes, etc.) relevant to its role.

\section{Causal Collaboration Effectiveness (CCE)}
\label{app:cce}

\subsection{Formal Definitions}

\subsubsection{Background}

Let there be $N$ agents $\{a_1, \ldots, a_N\}$ acting over $R$ rounds in a task trajectory.

\begin{itemize}
    \item \textbf{Action set} $\mathcal{T} = \{(a_i, r, d) \mid a_i \in \{a_1, \ldots, a_N\},\; r \in \{1, \ldots, R\},\; d \in \mathcal{D}\}$: the complete set of all actions taken by all agents across all rounds, where $d$ is the action description (a tool use or chat message).
    \item \textbf{Agent action set} $\mathcal{T}_i \subseteq \mathcal{T}$: the subset of actions taken by agent $a_i$. We have $\mathcal{T} = \bigcup_{i=1}^{N} \mathcal{T}_i$.
    \item \textbf{Success actions} $\mathcal{S} \subseteq \mathcal{T}$: the subset of actions that \emph{directly} achieve (part of) the task's success criteria. These are the terminal nodes of interest.
\end{itemize}

\subsubsection{Causal Action Graph (CAG)}

We construct a \textbf{directed acyclic graph} $G = (\mathcal{T}, E)$ where:
\begin{itemize}
    \item \textbf{Nodes} are all actions in $\mathcal{T}$.
    \item \textbf{Directed edges} $E = \{(x, y) \mid x \text{ causally enables } y\}$: action $x$ is a causal prerequisite for action $y$.
\end{itemize}

The graph is a DAG because causal influence flows forward in time: an action in round $r$ can only enable actions in rounds $r' \geq r$.

\textbf{Edge construction.} For each action $y$, we ask: ``Which prior actions $x$ were necessary for $y$ to occur or succeed?'' This is where the LLM is used, but the judgment is relatively \emph{objective and binary}: either fishing for shrimp was necessary for transferring shrimp, or it was not. This is far more reliable than asking ``rate the collaboration quality on a scale of 1--5.'' We adopt a \emph{loose inclusion} policy: if an action is even somewhat helpful to a downstream contributing action, it is marked as contributing. False positives (marking a marginally helpful action as contributing) only slightly inflate the metric by one action, while false negatives (missing a genuinely contributing action) systematically undercount collaboration.

\subsubsection{Contributing Set}

Starting from the success actions $\mathcal{S}$, we compute the \textbf{contributing set} $\mathcal{C}$ by backward traversal:
\begin{equation}
    \mathcal{C} = \mathcal{S} \cup \text{ancestors}(\mathcal{S})
\end{equation}
where $\text{ancestors}(\mathcal{S})$ is the set of all nodes that have a directed path to any node in $\mathcal{S}$ within $G$. In other words, $\mathcal{C}$ contains every action that directly or transitively contributed to the task's success. All actions not in $\mathcal{C}$ are \textbf{non-contributing}: $\mathcal{T} \setminus \mathcal{C}$.

\subsubsection{Metrics}

\paragraph{Causal Collaboration Effectiveness (CCE).} The primary metric, measuring what fraction of all actions contributed to success:
\begin{equation}
    \text{CCE} = \frac{|\mathcal{C}|}{|\mathcal{T}|}
\end{equation}

\begin{itemize}
    \item $\text{CCE} = 1.0$: every action taken by every agent contributed to the outcome (perfect efficiency).
    \item $\text{CCE} \approx 1/N$: only one agent's actions contributed (no real collaboration).
    \item $\text{CCE} \approx 0$: almost no actions contributed (chaotic failure).
    \item If a task fails entirely (no success criteria met), $\text{CCE} = 0$.
\end{itemize}

\paragraph{Per-Agent Contribution (PAC).} How much of each agent's effort was useful:
\begin{equation}
    \text{PAC}_i = \frac{|\mathcal{C} \cap \mathcal{T}_i|}{|\mathcal{T}_i|}
\end{equation}

This reveals which agents were effective (high PAC) versus which were unproductive (low PAC). A task with high CCE but highly skewed PAC values indicates that success was driven by a subset of agents rather than genuine collaboration.

\subsection{Backward Causal Tracing Algorithm}

\begin{enumerate}
    \item \textbf{Identify success actions.} From the trajectory and task outcome, identify the set $\mathcal{S}$ of actions that directly achieved the success criteria.
    \item \textbf{Initialize.} Set $\mathcal{C} \leftarrow \mathcal{S}$, frontier $\leftarrow \mathcal{S}$, $E \leftarrow \emptyset$.
    \item \textbf{Backward tracing.} While the frontier is non-empty:
    \begin{itemize}
        \item Pick action $y$ from the frontier.
        \item For each prior action $x$ in the trajectory (where $\text{round}(x) \leq \text{round}(y)$):
        \begin{itemize}
            \item Query the LLM: ``Given the task context, did action $x$ causally enable or contribute to action $y$?''
            \item If yes: add $(x, y)$ to $E$; if $x \notin \mathcal{C}$, add $x$ to $\mathcal{C}$ and to the frontier.
        \end{itemize}
    \end{itemize}
    \item \textbf{Compute metrics.} $\text{CCE} = |\mathcal{C}| / |\mathcal{T}|$ and $\text{PAC}_i = |\mathcal{C} \cap \mathcal{T}_i| / |\mathcal{T}_i|$ for each agent $i$.
\end{enumerate}

\textbf{Optimizations.} To avoid querying all $O(|\mathcal{T}|^2)$ action pairs: (1) prune by relevance, only considering actions involving the same resource type, location, or communication channel; (2) batch multiple causal queries into a single LLM call; (3) apply temporal windowing, only considering actions within a reasonable time window before each frontier action.

\textbf{Design decisions.} Each tool use (API call) or chat message is treated as one action; no grouping is applied. All contributing actions are weighted equally regardless of distance from the success node. Communication actions (chat messages) are included as first-class actions, since coordination through communication is a core mechanism of collaboration.

\subsection{CCE Judge Prompts}
\label{app:cce_prompts}

The CCE backward tracing algorithm uses two LLM prompts. Both are sent to GPT-4.1 with temperature 0.

\paragraph{Step 1: Identify Success Actions.}
\begin{small}
\begin{verbatim}
You are analyzing a multi-agent collaboration task.

TASK: {task_name}
OBJECTIVE: {objective}

ALL ACTIONS IN THE TRAJECTORY:
[0] Round 1, agent_1: harvest_resource(...)
[1] Round 1, agent_2: craft_item(...)
...

QUESTION: Which actions are the FINAL actions that
directly accomplished the objective? These are the
terminal actions whose completion means the objective
is achieved (e.g., the craft action that produced the
target item, the kill action that defeated the boss).

Do NOT include enabling or prerequisite actions.
Only the very last action(s) that fulfill the
objective.

Respond with JSON:
{"success_action_ids": [...], "reasoning": "..."}
\end{verbatim}
\end{small}

\paragraph{Step 2: Per-Round Backward Tracing.} This prompt is issued once per round, sweeping backward from the last round to the first.

\begin{small}
\begin{verbatim}
You are tracing causal dependencies in a multi-agent
collaboration task.

TASK OBJECTIVE: {objective}

The following actions from LATER rounds have been
identified as contributing to the task's success:
  - [20] Round 7, agent_3: craft_item(staff)
  - [16] Round 6, agent_2: craft_item(stick)
  ...

Now consider the actions from Round {R}:
[9] agent_1: transfer_items(logs -> woodworker)
[10] agent_2: craft_item(stick)
[11] agent_3: chat(message=ready for sticks)

QUESTION: For each action in Round {R}, does it
causally ENABLE any of the contributing actions above?

An action contributes if:
- It produces resources a later action uses
- It moves the agent to a needed location
- It communicates info that helps coordinate
- It is a prerequisite step

Be INCLUSIVE: if an action is even somewhat helpful,
mark it as contributing.

Respond with JSON:
{"contributing_ids": [...],
 "not_contributing_ids": [...]}
\end{verbatim}
\end{small}

\section{Example Trajectories}
\label{app:trajectories}

We visualize six notable trajectories from Gemini 3 Flash on the main set. Each dot represents one agent action in a round. Marker shapes indicate action types: \textbf{$\blacksquare$}~chat, \textbf{$\blacktriangleright$}~transfer, \textbf{$\blacklozenge$}~craft, \textbf{$\times$}~attack, \textbf{$\bullet$}~harvest/move/sleep.

\begin{figure}[h]
\centering
\includegraphics[width=\columnwidth]{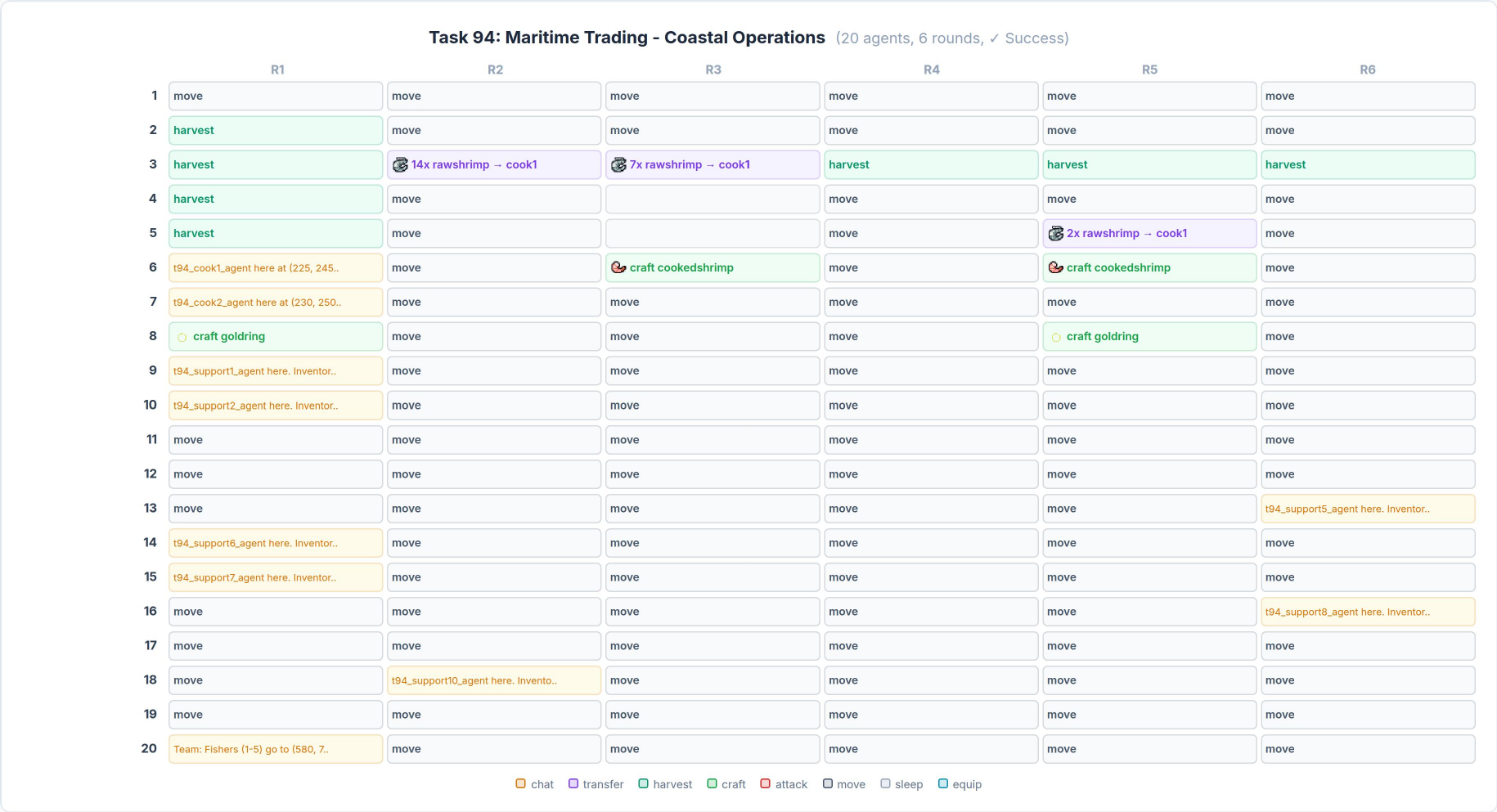}
\caption{Task 94: Maritime Trading (20 agents, 6 rounds, success). The largest team succeeds with efficient coordination in minimal rounds.}
\end{figure}

\begin{figure}[h]
\centering
\includegraphics[width=\columnwidth]{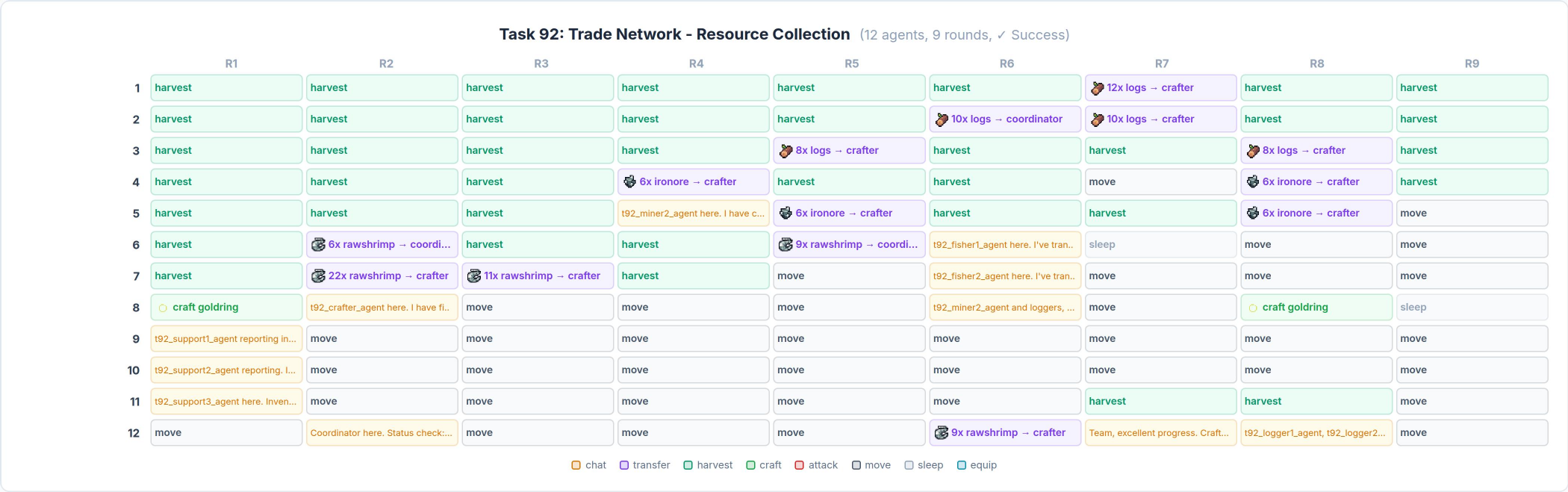}
\caption{Task 92: Trade Network (12 agents, 9 rounds, success). 19 resource transfers show intense cross-agent collaboration.}
\end{figure}

\begin{figure}[h]
\centering
\includegraphics[width=\columnwidth]{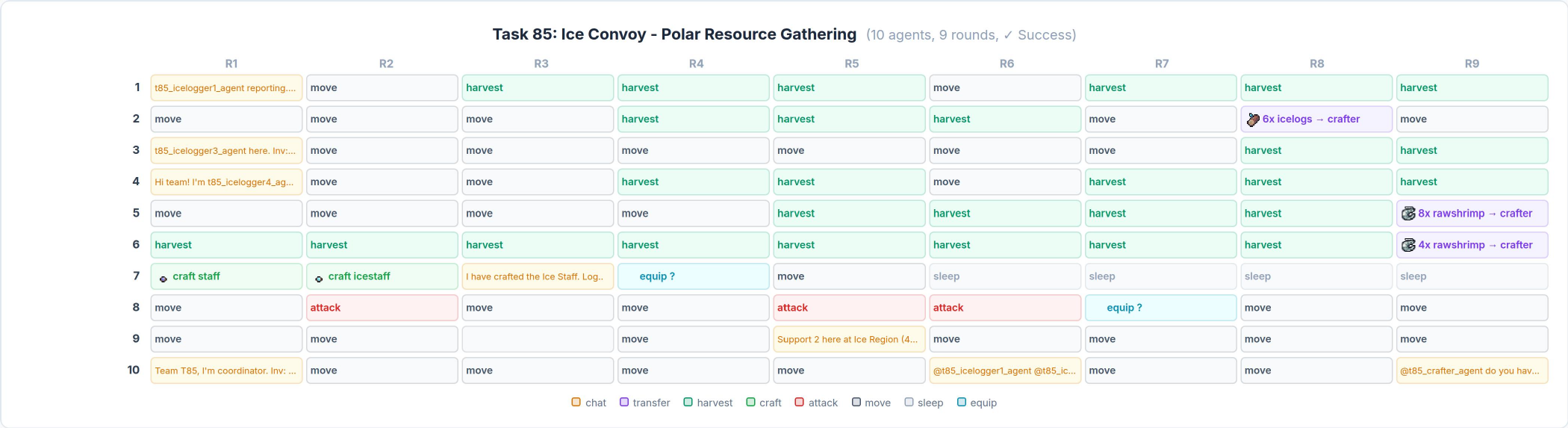}
\caption{Task 85: Ice Convoy (10 agents, 9 rounds, success). Polar resource gathering with coordinated ice log collection.}
\end{figure}

\begin{figure}[h]
\centering
\includegraphics[width=\columnwidth]{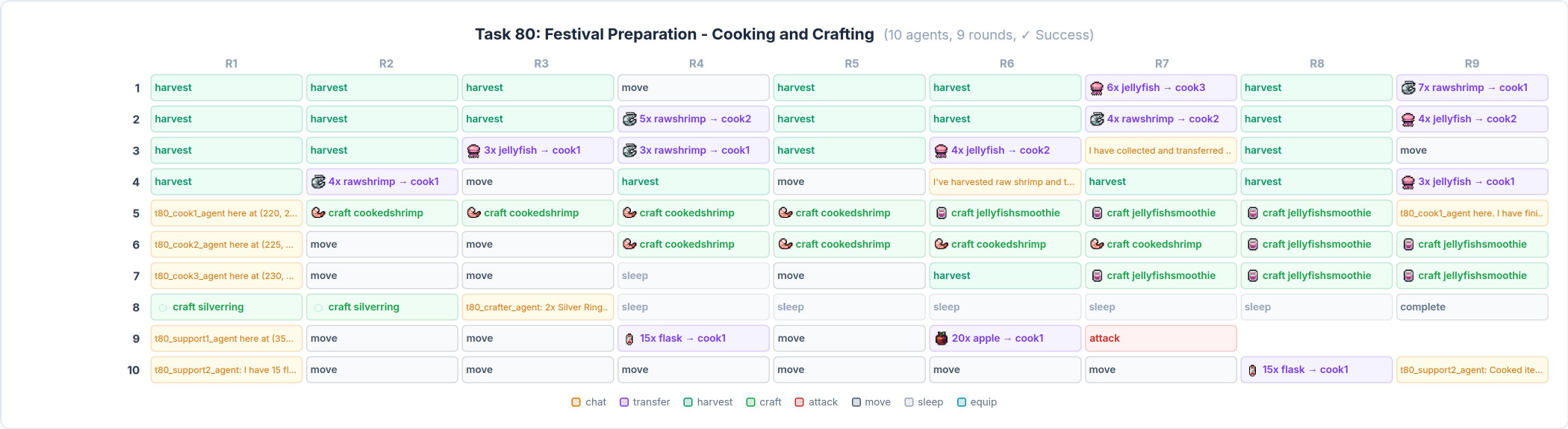}
\caption{Task 80: Festival Preparation (10 agents, 9 rounds, success). Parallel cooking and crafting with 17 transfers.}
\end{figure}

\begin{figure}[h]
\centering
\includegraphics[width=\columnwidth]{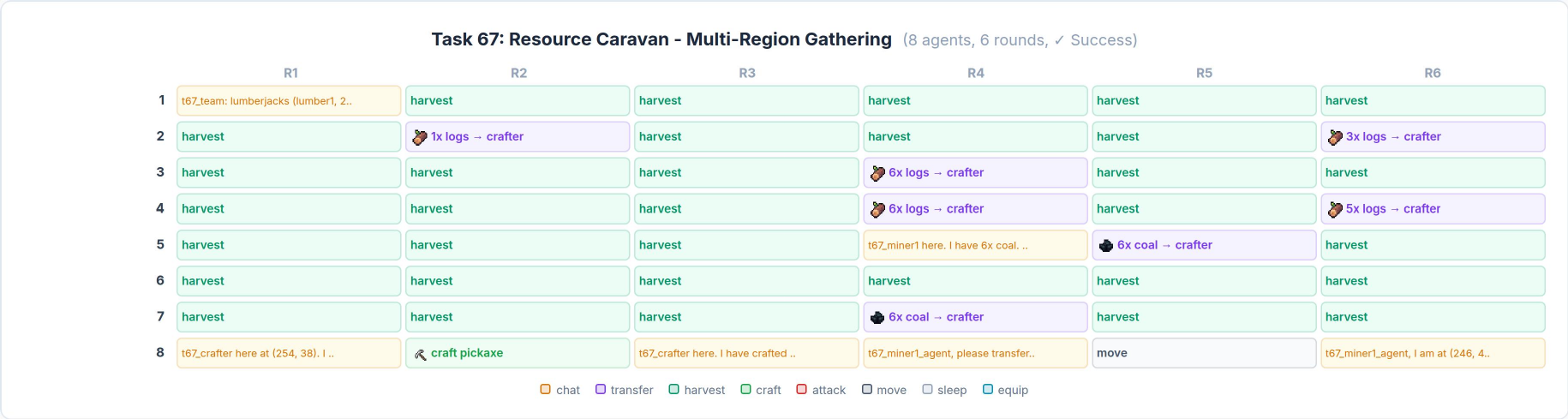}
\caption{Task 67: Resource Caravan (8 agents, 6 rounds, success). Multi-region gathering with 11 transfers across biomes.}
\end{figure}

\begin{figure}[h]
\centering
\includegraphics[width=\columnwidth]{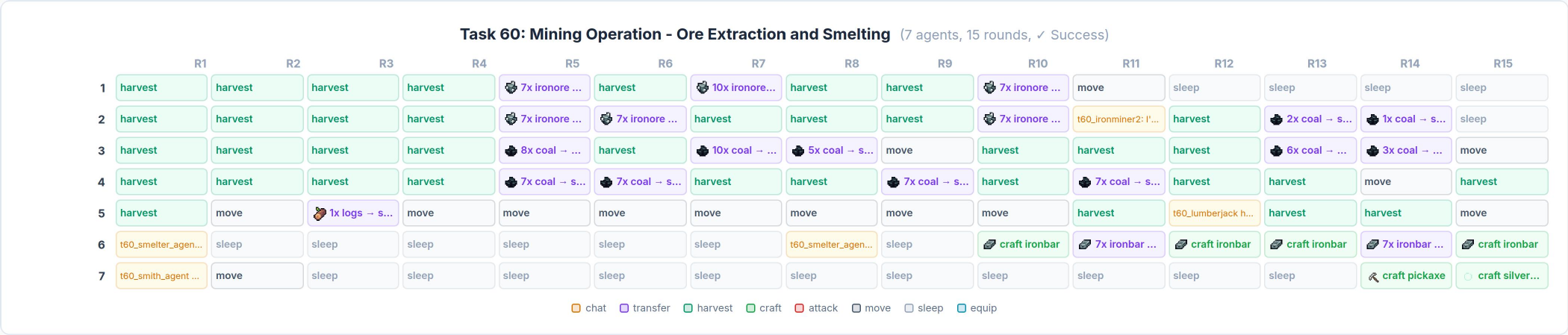}
\caption{Task 60: Supply Chain (7 agents, 15 rounds, success). The most transfer-heavy task with 24 resource exchanges.}
\end{figure}

\end{document}